\documentclass[twocolumn]{aastex62}

\received{27 October 2019}
\revised{16 June 2020}
\accepted{--}

\shorttitle{Black Hole Feedback Valve}
\shortauthors{Voit et al.}

\begin{document}

\title{\bf A Black Hole Feedback Valve in Massive Galaxies}
\author{G. Mark Voit}
\affiliation{Department of Physics and Astronomy,
                 Michigan State University,
                 East Lansing, MI 48824} 

\author{Greg L. Bryan}
\affiliation{Department of Astronomy,
                 Columbia University,
                 New York, NY} 
\affiliation{Center for Computational Astronomy,
                 Flatiron Institute,
                 New York, NY} 

\author{Deovrat Prasad}
\affiliation{Department of Physics and Astronomy,
                 Michigan State University,
                 East Lansing, MI 48824} 

\author{Rachel Frisbie}
\affiliation{Department of Physics and Astronomy,
                 Michigan State University,
                 East Lansing, MI 48824} 
                 
\author{Yuan Li}
 \affiliation{Center for Computational Astronomy,
                 Flatiron Institute,
                 New York, NY} 
\affiliation{Department of Astronomy,
                 University of California, Berkeley,
                 Berkeley, CA}

\author{Megan Donahue}
\affiliation{Department of Physics and Astronomy,
                 Michigan State University,
                 East Lansing, MI 48824} 

\author{Brian W. O'Shea}
\affiliation{Department of Physics and Astronomy,
                 Michigan State University,
                 East Lansing, MI 48824} 
\affiliation{Department of Computational Mathematics, Science, and Engineering,
                 Michigan State University,
                 East Lansing, MI} 
\affiliation{National Superconducting Cyclotron Laboratory,
                 Michigan State University,
                 East Lansing, MI} 
                 
\author{Ming Sun}  
\affiliation{Department of Physics and Astronomy,
                 The University of Alabama in Huntsville,
                 Huntsville, AL} 

\author{Norbert Werner}  
\affiliation{MTA-E\"otv\"os University Lend\"ulet Hot Universe Research Group, 
	        P\'azm\'any P\'eter s\'et\'any 1/A, Budapest, 1117, 
	        Hungary} 
\affiliation{Department of Theoretical Physics and Astrophysics, Faculty of Science, Masaryk University,
	        Kotl\'a\v{r}sk\'a 2, Brno, 611 37, Czech Republic} 
\affiliation{School of Science, Hiroshima University, 1-3-1 Kagamiyama, Higashi-Hiroshima 739-8526,
 	        Japan}

\begin{abstract}
Star formation in the universe's most massive galaxies proceeds furiously early in time but then nearly ceases.  Plenty of hot gas remains available but does not cool and condense into star-forming clouds.  Active galactic nuclei (AGNs) release enough energy to inhibit cooling of the hot gas, but energetic arguments alone do not explain why quenching of star formation is most effective in high-mass galaxies.  In fact, optical observations show that quenching is more closely related to a galaxy's central stellar velocity dispersion ($\sigma_v$) than to any other characteristic.  Here we show that high $\sigma_v$ is critical to quenching because a deep central potential well maximizes the efficacy of AGN feedback.  In order to remain quenched, a galaxy must continually sweep out the gas ejected from its aging stars.  Supernova heating can accomplish this task as long as the AGN sufficiently reduces the gas pressure of the surrounding circumgalactic medium (CGM).  We find that CGM pressure acts as the control knob on a valve that regulates AGN feedback and suggest that feedback power self-adjusts so that it suffices to lift the CGM out of the galaxy's potential well.  Supernova heating then drives a galactic outflow that remains homogeneous if $\sigma_v \gtrsim 240 \, {\rm km \, s^{-1}}$.  The AGN feedback can effectively quench galaxies with a comparable velocity dispersion, but feedback in galaxies with a much lower velocity dispersion tends to result in convective circulation and accumulation of multiphase gas within the galaxy.
\end{abstract}



\section{Introduction}
\setcounter{footnote}{0}
\label{sec-intro}

The surest way to quench star formation in a galaxy is to rid it of molecular gas.  That end can be accomplished either gradually, by turning molecular gas into stars faster than it can accumulate, or abruptly, by destroying or ejecting all of the galaxy's molecular clouds.  However, star formation will eventually resume unless the galaxy can prevent more molecular gas from accumulating.  Three different gas sources must therefore be prevented from supplying cold gas.
\begin{enumerate}
\item \textit{Cold streams.} Accretion of cold gas along cosmological dark matter filaments can potentially feed star formation \citep[e.g.,][]{Keres_2005MNRAS.363....2K,Keres_2009MNRAS.396.2332K,Dekel_2009Natur.457..451D}.  If not disrupted before reaching the bottom of the local potential well, those cold accretion streams will enter the central galaxy.   Most current models of galaxy evolution therefore posit that quiescent central galaxies have hot gaseous halos that disrupt cold streams.  They also posit that quiescent satellite galaxies orbiting the central one cannot access cold streams because of their displacement from the center.
\item \textit{Cooling flows.}  Even if cold streams are disrupted by a hot halo, radiative cooling of the densest gas in that hot halo can still supply cold gas to the central galaxy \citep[e.g.,][]{WhiteFrenk1991ApJ...379...52W,Fabian94}.  Most current models of galaxy formation therefore posit that accretion of cooling gas onto a central supermassive black hole releases enough energy to offset most of the radiative losses, thereby limiting the supply of cold gas to the central galaxy \cite[e.g.,][]{mn07,McNamaraNulsen2012NJPh...14e5023M,Werner_2019SSRv..215....5W}.
\item \textit{Stellar mass loss.}  The third gas source is the aging stellar population of a quiescent galaxy \citep[e.g.,][]{MathewsBrighenti2003ARAA..41..191M,LeitnerKravtsov_2011ApJ...734...48L,VoitDonahue2011ApJ...738L..24V}.  As dying stars shed their surplus gas, it accumulates in the galaxy, where it is heated by exploding white dwarfs (SNe Ia).  Supernova heating is energetically capable of sweeping ejected stellar gas out of a galaxy, but the pressure of the galaxy's circumgalactic medium (CGM) limits the rate at which it can do so.  That is because the confining CGM pressure determines the outflow's gas density and its radiative losses.  If the CGM pressure is too great, supernovae cannot sweep out the ejected stellar gas.
\end{enumerate}

This paper analyzes the three-way coupling that can occur between supernova sweeping of stellar ejecta, the confining CGM pressure, and bipolar kinetic feedback fueled by accretion of cooling gas onto the central black hole.  Together, they make a valve that regulates fueling of the active galactic nucleus (AGN).  The ideas presented build upon an enormous literature that is impossible to adequately summarize here, and so we will restrict ourselves to pointing out some highlights.  

\citet{MathewsBaker1971ApJ...170..241M}, in a seminal and insightful paper on galactic winds, laid a foundation that remains strong nearly half a century later.  Interestingly, their pursuit of outflow solutions for supernova-driven galactic winds was equally motivated by both the absence of cold interstellar gas and the presence of nonthermal radio sources in massive elliptical galaxies.  \citet{MathewsBrighenti2003ARAA..41..191M} have extensively reviewed the line of research that followed.

A couple of decades later, \citet{TaborBinney1993MNRAS.263..323T} and \cite{BinneyTabor_1995MNRAS.276..663B} focused more closely on how cooling-flow accretion onto a central black hole in a massive elliptical can limit the overall condensation rate.  They recognized that the radial profiles of pressure, density, and temperature inferred from X-ray observations resemble galactic outflows near the onset of a cooling catastrophe that supernova heating cannot prevent.  While supernova heating can drive gas out of a massive elliptical galaxy, it cannot unbind that gas from the galaxy's halo \citep[e.g.,][]{MathewsLoewenstein_1986ApJ...306L...7M,LoewensteinMathews_1987ApJ...319..614L,DavidFormanJones_1990ApJ...359...29D}.  Without additional heating, ejected stellar mass therefore collects in the CGM and raises the confining pressure.  Eventually, the central density of the confined outflow increases enough for radiative cooling to exceed supernova heating, and cooling gas starts to flow toward the origin.  Binney \& Tabor proposed that the central cooling flow should then fuel accretion onto a supermassive black hole, producing bipolar jets that heat and expand the surrounding medium, raising its entropy and temporarily alleviating the cooling catastrophe.

Around the same time, \citet{Ciotti_1991ApJ...376..380C} launched a series of increasingly sophisticated simulations, at first to explore the evolution of supernova-heated outflows and later to explore the coupling between those outflows and sporadic black hole outbursts \citep{CiottiOstriker_2001ApJ...551..131C,CiottiOstriker_2007ApJ...665.1038C,Ciotti_2010ApJ...717..708C,Ciotti_2017ApJ...835...15C}.  Those models demonstrated the importance of both kinetic feedback \citep{OstrikerCiotti_2010ApJ...722..642O,Choi_2015MNRAS.449.4105C}
and the circumgalactic environment \citep{Shin_2012ApJ...745...13S} to the evolution of hot gas in massive ellipticals, but they did not directly explore the effects of AGN feedback on the confining CGM pressure.  During much of the time, the simulated AGN power is relatively low, allowing radiative cooling to reduce the central entropy and raise the central gas density until a cooling catastrophe begins.  The rapid rise in cooling then fuels a strong burst of AGN feedback, lasting several tens of Myr, that boosts the central entropy and cooling time while lowering the central gas density. 

More recently, three-dimensional simulations have vividly illustrated how kinetic feedback limits cooling and condensation of ambient galactic gas and why it is more effective than purely thermal feedback \citep[e.g.,][]{Gaspari+2011MNRAS.411..349G,Gaspari+2011MNRAS.415.1549G,Gaspari+2012ApJ...746...94G,LiBryan2014ApJ...789...54L,Li_2015ApJ...811...73L,Dubois+2012MNRAS.420.2662D,
Prasad_2015ApJ...811..108P,Prasad_2017MNRAS.471.1531P,YangReynolds_2016ApJ...829...90Y,Meece_2017ApJ...841..133M,Beckmann_2019}.
 Strong, narrow jets can drill through the ambient medium within a few kpc of the AGN without significantly disturbing much of it, and that enables the jets to transport most of their kinetic energy tens of kpc from the origin before thermalizing it in the CGM \citep[see also][]{Soker_2016NewAR..75....1S}.  This transport mechanism distributes thermal energy over a large volume without producing excessive convection, which destabilizes the ambient medium and results in overproduction of multiphase gas \citep{Meece_2017ApJ...841..133M,Voit_2017_BigPaper}

Here we build upon those previous efforts by applying recent insights into the thermal stability of the CGM.  Our objective is to determine the conditions that allow supernova heating to sweep ejected stellar mass out of a massive elliptical galaxy in an outflow that remains stable to multiphase condensation \citep[see][hereafter V15, for a preliminary analysis]{Voit+2015ApJ...803L..21V}.  In order for the ambient medium to remain homogeneous, its ratio of radiative cooling time ($t_{\rm cool}$) to freefall time ($t_{\rm ff}$) must satisfy the condition $t_{\rm cool} / t_{\rm ff} \gtrsim 10$ (see \S \ref{sec-ThermalStability}).   Also, supernova heating must exceed radiative cooling within most of the galaxy's volume, requiring the ambient gas density to decline with radius at least as rapidly as $\propto r^{-1}$.  In that case, radiative cooling per unit volume declines at least as rapidly as supernova heating, which is proportional to the stellar mass density (approximately $\rho_* \propto r^{-2}$).  These conditions cannot be satisfied unless something other than supernova heating lowers the pressure of the CGM surrounding the galaxy.  We therefore assume that powerful but sporadic AGN outbursts clear the way for supernova sweeping by lowering the confining CGM pressure.  Those outbursts occur whenever cooling at small radii supplies the central black hole with an unusually large amount of fuel.  As a consequence, the CGM pressure ultimately governs the maximum black hole fueling rate because it determines the mass inflow rate of the inner cooling flow.  

The most far-reaching result of this paper is that this critical gas density profile ($\rho \propto r^{-1}$) corresponds to a critical stellar velocity dispersion ($\sigma_v \approx 240 \, {\rm km \, s^{-1}}$), above which coupling of AGN feedback to CGM pressure can keep star formation permanently quenched.  Nearly complete quenching can happen because the ambient medium then satisfies the conditions for supernova heating to drive a steady, homogeneous flow of ejected stellar gas out of the galaxy.   

Readers interested in the basic physical picture should start with Section \ref{sec-BasicPicture}, which shows that the critical density slope is determined by a critical entropy profile slope that depends almost exclusively on the ratio of the specific energy of stellar ejecta to the square of the gravitational potential's circular velocity.  It therefore depends on the quotient of the specific supernova heating rate and specific stellar mass-loss rate, which both evolve with time.  Section \ref{sec-SteadyFlow} presents a more formal treatment, adding rigor to the analytical estimates of \S \ref{sec-BasicPicture} by comparing them with numerical integrations of the steady flow equations.  Observational support for the model can be found in \S \ref{sec-ObsComparison}, which validates the numerical solutions through comparisons with X-ray observations of massive elliptical galaxies and infers the role of intermittent AGN feedback from the discrepancies.  Section \ref{sec-Implications} discusses the resulting implications and predictions for the quenching of star formation, focusing in particular on why quenching should depend primarily on central stellar velocity dispersion and secondarily on halo mass.  Section \ref{sec-Summary} summarizes the paper.

\section{The Basic Picture}
\label{sec-BasicPicture}

This section outlines the black hole feedback valve model by presenting a series of simple analytical estimates intended to make the overall physical picture intuitively clear.  It begins by modeling the hot ambient medium as a series of concentric spherical shells, each with a constant temperature, so that gas density depends only on specific entropy.  It then shows how the structure of the resulting galactic atmosphere depends on the entropy profile slope determined by the average specific energy of stellar ejecta.  It explains how the radius $r_{\rm eq}$ at which local supernova heating equals local radiative cooling is determined by an outer pressure boundary condition, and it demonstrates that supernova heating cannot sweep ejected stellar gas out of the galaxy unless AGN feedback reduces the confining CGM pressure to a small fraction of its cosmological value.  The section concludes with an assessment of the conditions under which the outflow is stable to multiphase condensation and some estimates of the AGN feedback power and momentum flux required to make the basic picture self-consistent.

\subsection{Piecewise Isothermal Atmosphere}

A galaxy's ambient atmosphere can be approximated with a piecewise isothermal hydrostatic model as long as its velocity field is sufficiently subsonic.   Quantifying specific entropy in terms of the entropy index $K \equiv kT n_e^{-2/3}$, where $T$ is the gas temperature and $n_e$ is the electron density, leads to 
\begin{equation}
  \frac {d \ln P} {d \ln r} = - \frac {3} {2} \frac {d \ln K} {d \ln r} = - \frac {\mu m_p v_c^2} {kT}
\end{equation}
within each hydrostatic, isothermal shell of gas in a gravitational potential with circular velocity $v_c$.  The logarithmic entropy slope in each shell is
\begin{equation}
  \alpha_K \equiv \frac {d \ln K} {d \ln r} = \frac {2} {3} \frac {\mu m_p v_c^2} {kT}
  \label{eq-alpha_K-T}
  \; \; , 
\end{equation}
giving $K \propto r^{\alpha_K}$ and $n_e \propto r^{-3 \alpha_K / 2}$ for an isothermal potential in which $v_c$ is constant with radius.  Within each shell, the constant of proportionality relating hydrostatic temperature to $v_c^2$ is determined by the entropy slope $\alpha_K$ of the galactic atmosphere in that shell.  A galactic atmosphere in which $T$ is a slowly changing function of radius can therefore be approximated with a set of thick concentric isothermal shells in which $T$ and $\alpha_K$ are related through equation (\ref{eq-alpha_K-T}).

Cosmological structure formation tends to produce dark matter halos similar to singular isothermal spheres in which gas density is approximately proportional to dark matter density.  Hydrostatic gas in an idealized cosmological halo can therefore be approximated by choosing $\alpha_K = 4/3$, $n_e \propto r^{-2}$, and $T = T_\phi$, with $T_\phi \equiv \mu m_p v_c^2 / 2 k \approx {\rm const.}$,\footnote{A more realistic approximation would set $\alpha_K = 1.1$ and assume a non-isothermal NFW potential \citep{nfw97}.} but radiative cooling generally alters the structure of the inner regions.  Without a nongravitational heat source, radiative cooling in an isothermal potential produces a central cooling flow that has $\alpha_K = 1$, $n_e \propto r^{-3/2}$, and $T = 4 T_\phi / 3$ \citep[e.g.,][]{Voit_2017_BigPaper,Stern_CoolingFlows_2019MNRAS.488.2549S}.  However, cooling of gas near the center of the potential well can also result in star formation followed by supernova explosions that generate heat.

\subsection{Supernova Sweeping}
\label{sec-Sweeping}

Long after star formation has ceased, an aging stellar population continues to supply mass and energy to the ambient medium through normal stellar mass loss, planetary nebulae, and SNe Ia.  In order for star formation to remain quenched, the galaxy must sweep out the gas shed by stars as quickly as it accumulates through some combination of supernova heating and AGN feedback \citep[e.g.,][V15]{MathewsBrighenti2003ARAA..41..191M,David_2006ApJ...653..207D,VoitDonahue2011ApJ...738L..24V}.  This section shows that the entropy slope of the resulting outflow in regions where supernova heating locally exceeds radiative cooling depends primarily on the ratio $\epsilon_* / v_c^2$ relating the specific thermal energy $\epsilon_*$ of ejected stellar gas to the depth of the galactic potential well.

The radial structure of such a steady spherical flow depends on its Bernoulli specific energy $\epsilon$, defined by 
\begin{equation}
  \epsilon (r)  \equiv \frac {v_r^2} {2} + \phi (r)  + \frac {5} {2} \frac {kT} {\mu m_p}
  \; \; ,
\end{equation}
where $v_r$ is the flow's radial velocity, $\phi (r)$ is the gravitational potential, and the enthalpy term ($5kT/2 \mu m_p$) accounts for both the specific thermal energy and the $PdV$ work done by the flow.  If radiative losses are negligible, the steady-state flow of Bernoulli energy through radius $r$ equals the integrated rate of energy input into the gas within the volume bounded by $r$.  Three energy sources contribute. 
\begin{enumerate}
\item \textit{Supernova energy.} Dividing the rate of stellar heat input by the rate of stellar mass loss gives the mean specific energy $\epsilon_*$ of gas coming from stars. 
\item  \textit{Orbital energy.} The equilibrium isotropic velocity dispersion of a singular isothermal sphere of stars is $\sigma_v = v_c / \sqrt{2}$, and so thermalization of the orbital kinetic energy associated with stellar mass loss adds a specific energy $3 v_c^2 / 4$.
\item \textit{Gravitational potential energy.} The integrated stellar mass loss within $r$ adds a mean (mass-weighted) specific potential energy $\bar{\phi}(r)$.
\end{enumerate}  
The Bernoulli specific energy of outflowing gas at radius $r$ is therefore
\begin{equation}
  \epsilon (r) = \epsilon_* + \frac {3} {4} v_c^2 + \bar{\phi}(r)
\end{equation}
if there are no radiative losses.

As long as the outflow remains highly subsonic, its specific kinetic energy can be neglected, and its temperature is
\begin{equation}
  kT \approx \frac {2} {5} \mu m_p \left[ \epsilon_* + \frac {3} {4} v_c^2 - (\phi - \bar{\phi} ) \right]
  \; \; .
\end{equation}
In a singular isothermal sphere dominated by stellar mass, one finds $(\phi - \bar{\phi}) = v_c^2$, because $\phi(r) = v_c^2 \ln (r/r_\phi)$ and $\bar{\phi}(r) =  v_c^2 [\ln (r/r_\phi) - 1 ]$.  According to equations (2) and (5), the power-law entropy slope of the outflow in that region is then
\begin{equation}
  \alpha_K \approx \frac {5} {3} \left( \frac {\epsilon_*} {v_c^2} - \frac {1} {4} \right)^{-1}
  \; \; ,
   \label{eq-alpha_K_vc}
\end{equation}
demonstrating that the atmosphere's structure depends almost entirely on the ratio $\epsilon_* / v_c^2$ \citep[see also][V15]{SharmaNath_2013ApJ...763...17S}.

If the gravitational potential is negligible compared to the specific energy of stellar mass loss (i.e. $v_c^2 \ll \epsilon_*$), little work is required to drive the outflow.  In that case, the region where stars are adding mass and energy is nearly isobaric and isentropic, with $kT \approx (2/5) \mu m_p \epsilon_*$.  At the other extreme, as $v_c^2$ approaches $4 \epsilon_*$, the outflow's pressure and entropy gradients become extremely steep because the specific energy of stellar mass loss becomes incapable of driving a steady outflow through the isothermal potential.  However, the case most relevant to quenching of isolated massive galaxies is $v_c^2 \approx \epsilon_* / 2$, in which the work necessary to drive the outflow is comparable to the supernova heat input, resulting in $\alpha_K \sim 1$.

\subsection{A Critical Entropy Slope}
\label{sec-CriticalSlope}

The primary predictions of this paper follow from the fact that galactic outflows with $\alpha_K > 2/3$ should fundamentally differ from those with $\alpha_K < 2/3$.  At the crossover (i.e., $\alpha_K = 2/3$), the electron density profile of isothermal gas is $n_e \propto r^{-1}$, meaning that radiative cooling per unit volume is $\propto r^{-2}$.  In that configuration, the ratio of radiative cooling per unit volume to stellar heating per unit volume is constant with radius within the region where the stellar mass density is $\propto r^{-2}$.   The ratio of stellar heating to radiative cooling consequently decreases with radius in galaxies with a shallow ambient entropy slope ($\alpha_K < 2/3$) and rises with radius in galaxies with a steep ambient entropy slope ($\alpha_K > 2/3$).

Galaxies in which $\alpha_K > 2/3$ can therefore remain in a steady state consisting of a cooling flow encompassed within a supernova-heated outflow.  Those two flows diverge from a stagnation radius near $r_{\rm eq}$ representing both the outer boundary of the cooling flow and the inner boundary of the outflow.  Mass shed by stars at $\lesssim r_{\rm eq}$ flows inward, while mass shed by stars at $\gtrsim r_{\rm eq}$ flows outward.

Another special feature of $\alpha_K = 2/3$ is that the ratio of cooling time to freefall time is then constant in an isothermal potential.  The $t_{\rm cool} / t_{\rm ff}$ ratio of the ambient medium in a galaxy with $\alpha_K > 2/3$ consequently increases with radius, implying that the region of the ambient medium most susceptible to multiphase condensation is at small radii.  One therefore expects the inner cooling flow to produce cold clouds primarily in the vicinity of the central black hole, potentially supercharging the feedback output fueled by chaotic cold accretion 
\citep{Gaspari+2013MNRAS.432.3401G,Gaspari_2015A&A...579A..62G,Gaspari_2017MNRAS.466..677G}, without supplying molecular gas for star formation at larger radii.  In contrast, the ambient medium in a galaxy with $\alpha_K < 2/3$ is more susceptible to multiphase condensation at large radii than at small radii.

\begin{figure*}[t]
\begin{center}
\includegraphics[width=6.0in, trim = 0.0in 0.2in 0.0in 0.0in]{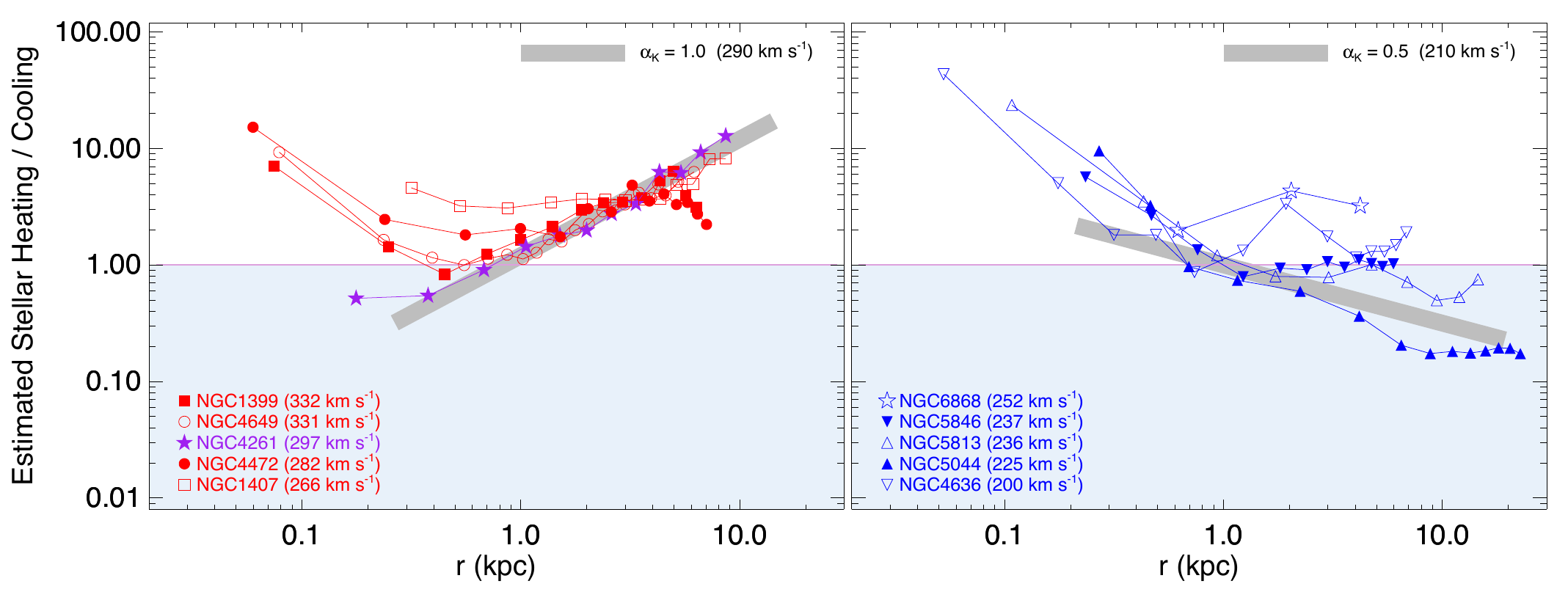} \\
\end{center}
\caption{ \footnotesize 
Estimated ratio of stellar heating per unit volume to radiative cooling per unit volume plotted as a function of radius for 10 massive elliptical galaxies, including five with no extended multiphase gas outside the central $\sim 1$~kpc (left panel) and five that have extended multiphase gas (right panel).  Radiative cooling rates are based on {\em Chandra} observations by \citet{Werner+2012MNRAS.425.2731W,Werner+2014MNRAS.439.2291W} and assume solar metallicity.  Stellar heating estimates assume that the stellar mass is distributed like a singular isothermal sphere with the central velocity dispersion given by Hyperleda (shown in parentheses), resulting in a heating rate per unit volume $(\epsilon_* + 3 \sigma_v^2 / 2) \rho_* / t_*$, given $\epsilon_* = 2 \, {\rm keV} / \mu m_p$ and $t_* = 200$~Gyr.  Blue shading shows where radiative cooling exceeds stellar heating.  Thick gray lines show the general trend predicted by equation (\ref{eq-alpha_K_vc}) for the velocity dispersions listed in the labels.
\vspace*{1em}
\label{Werner_equality_red_blue}}
\end{figure*}

Solving equation (\ref{eq-alpha_K_vc}) for the circular velocity corresponding to the critical entropy slope ($\alpha_K = 2/3$) gives
\begin{equation}
  v_c \approx \left( \frac {4 \epsilon_*} {11}  \right)^{1/2} 
        \approx  \; 340 \, {\rm km \, s^{-1}} \left( \frac {\mu m_p \epsilon_*} {2 \, {\rm keV}} \right)^{1/2}
        \; \; .
\end{equation}
The same result written in terms of an isotropic isothermal velocity dispersion is 
\begin{equation}
  \sigma_v \approx \left( \frac {2 \epsilon_*} {11}  \right)^{1/2}
                 \approx  240 \, {\rm km \, s^{-1}} \left( \frac {\mu m_p \epsilon_*} {2 \, {\rm keV}} \right)^{1/2}
        \; \; .
\end{equation}
Observations of the specific SN Ia rate from a stellar population with an age $\sim 10$~Gyr indicate that it is $\sim 3 \times 10^{-14} \, {\rm SN \, yr^{-1}} \, M_\odot^{-1}$ in massive elliptical galaxies \citep{Maoz2012MNRAS.426.3282M,FriedmannMaoz_2018MNRAS.479.3563F}.  Multiplying that rate by $10^{51}$ erg per SN Ia {\citep{Shen_2018ApJ...854...52S} and dividing the result by the specific stellar mass-loss rate ($t_*^{-1}$) gives \begin{equation}
  \epsilon_* \approx \frac {2 \, {\rm keV}} {\mu m_p} \cdot \frac {t_*} {200 \, {\rm Gyr}}
  \; \; .
\end{equation}
The 200~Gyr timescale chosen for scaling $t_*$ corresponds to 0.5\% of the stellar mass per Gyr and is broadly consistent with a stellar population age of $\sim 10$~Gyr but depends in detail on the stellar initial mass function \citep[IMF; e.g.,][]{LeitnerKravtsov_2011ApJ...734...48L}.  The specific SN Ia rate and specific stellar mass-loss rates are both time-dependent and somewhat uncertain, meaning that the critical velocity dispersion predicted by the model is also time-dependent and somewhat uncertain.\footnote{The inferred value of $\epsilon_*$ is more robust to assumptions about the IMF than either the specific stellar mass-loss rate or the specific SN Ia rate, as long as the same IMF is assumed.  Here the quoted SN Ia rate is relative to the stellar mass of the initial stellar population, which was assumed to have a \citet{Kroupa_2001MNRAS.322..231K} IMF.  For a similar IMF, the fits of \citet{LeitnerKravtsov_2011ApJ...734...48L} give $t_* = 217 \, {\rm Gyr}$, relative to the initial stellar mass, at a stellar population age of 10~Gyr.}   However, the ratio of SN Ia heating to stellar mass loss changes relatively slowly with time \citep[e.g.,][]{Ciotti_1991ApJ...376..380C}, meaning that the time dependence of the critical velocity dispersion should also be rather mild (but see \S \ref{sec-Evolution}).  

\subsection{Heating/Cooling Equality}
\label{sec-HC_Equality}

Radiative cooling per unit volume equals supernova heating per unit volume at any radius at which
\begin{equation}
  \left( \epsilon_* + \frac {3} {2} \sigma_v^2 \right)  \frac {\rho_*} {t_*} = n_e n_p \Lambda(T)
    \; \; ,
    \label{eq-HC_equality}
\end{equation}
where $\rho_*$ is the stellar mass density and $\Lambda(T)$ is the usual radiative cooling function, defined with respect to the proton density $n_p$.  Solving for $P$ gives the pressure profile \begin{equation}
  P_{\rm eq} (r)  \equiv \left[ \left( \epsilon_* + \frac {3} {2} \sigma_v^2 \right)
                          \left( \frac {n^2} {n_e n_p}  \right) \frac {\rho_*} {t_* \Lambda (T)} \right]^{1/2}
                          kT
  	\label{eq-P_eq} \\
\end{equation}
along which radiative cooling would equal supernova heating, given the temperature $T$. For $\sigma_v \approx 240 \, {\rm km \, s^{-1}}$, supernova sweeping without radiative cooling gives $\alpha_K \approx 2/3$ and $kT \approx 0.75$~keV.  The critical profiles of pressure, electron density, and entropy in solar metallicity gas near this temperature are
\begin{eqnarray}
  P_{\rm eq} (r) & \, \approx \, & ( 1.4 \times 10^{-10} \, {\rm erg \, cm^{-3}})
  				\, \sigma_{240}^3  \,  r_{\rm kpc}^{-1}  \\
  n_{e,{\rm eq}} (r) & \, \approx \,  &  ( 0.06 \, {\rm cm^{-3}} ) \, \sigma_{240} \, r_{\rm kpc}^{-1} \\ 
  K_{\rm eq} (r) & \, \approx \,  &  ( 5 \,  \, {\rm keV \, cm^2}) \, \sigma_{240}^{4/3}  \, r_{\rm kpc}^{2/3} 
       \; \; , 
\end{eqnarray}
given $r_{\rm kpc} \equiv r/(1 \, {\rm kpc})$ and $\sigma_{240} \equiv \sigma_v / 240 \, {\rm km \, s^{-1}}$, an isothermal stellar mass distribution ($\rho_* = \sigma_v^2 / 2 \pi G r^2$), and the fiducial values $\mu m_p \epsilon_* \approx 2$~keV and $t_* \approx 200$~Gyr, if the weak dependence of $\Lambda(T)$ on $\sigma_v$ is ignored. 

Figure~\ref{Werner_equality_red_blue} shows how the ratio of estimated stellar heating to radiative cooling depends on radius for 10 massive ellipticals with high-quality {\em Chandra} X-ray observations from \citet{Werner+2012MNRAS.425.2731W,Werner+2014MNRAS.439.2291W}.  The interstellar metallicity in each of them is observed to be approximately solar, and so we have computed their radiative cooling rates assuming solar metallicity throughout the paper.\footnote{This assumption is motivated by abundance observations and ignores the fact that the mean iron abundance of the gas ejected from old stars should be several times solar, given the observed SN Ia rate.  Consequently, the iron coming from recent SN Ia must not be well mixed with the gas coming from the rest of the stars, but the reason remains mysterious.  (See V15 for a brief discussion.)}   Our estimator for stellar heating per unit volume is the left-hand side of equation (\ref{eq-HC_equality}), assuming $\epsilon_* = 2 \, {\rm keV} / \mu m_p$, $t_* = 200$~Gyr, and $\rho_* = \sigma_v^2 / 2 \pi G r^2$, with $\sigma_v$ for each galaxy from Hyperleda.  It overestimates stellar heating at radii beyond where the stellar mass density starts to decline more steeply than $\rho_* \propto r^{-2}$, but for now, we are most interested in the range $0.5 \, {\rm kpc} \lesssim r \lesssim 2 \, {\rm kpc} $, where all of the estimated heating/cooling ratios are of order unity.

\begin{figure*}[t]
\begin{center}
\includegraphics[width=7.0in, trim = 0.0in 0.2in 0.0in 0.0in]{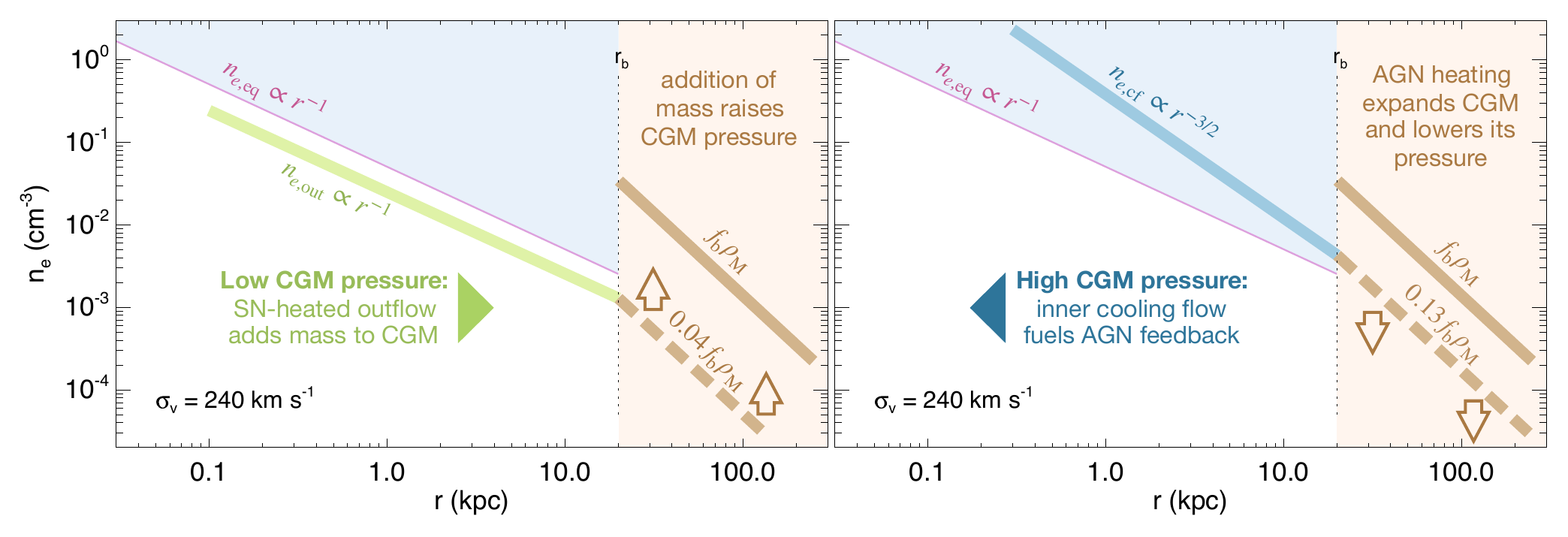} \\
\end{center}
\caption{ \footnotesize 
Schematic relationship between the flow pattern of ejected stellar gas and the boundary pressure around a galaxy with the critical velocity dispersion $\sigma_v \approx 240 \, {\rm km \, s^{-1}}$.  In the left panel, the CGM electron density (thick dashed tan line) at the boundary radius $r_{\rm b}$ is 4\% of the cosmological expectation (thick solid tan line).  Consequently, the electron density and pressure at $r_{\rm b}$ are lower than the level at which radiative cooling would equal stellar heating ($n_{e,{\rm eq}}$, thin violet line).  Stellar heating therefore drives an outflow with a density profile ($n_{e,{\rm out}}$, thick green line) that does not cross into the shaded region where cooling exceeds heating.  However, the outflow adds gas to the CGM and raises its pressure, leading to the situation in the right panel.  Here the electron density at $r_{\rm b}$ is 13\% of the cosmological expectation.  Radiative cooling therefore exceeds stellar heating at $r_{\rm b}$, causing an inflow ($n_{e,{\rm cf}}$, thick blue line) that fuels AGN feedback.  If thermalization of the resulting feedback energy in the CGM can add an amount of heat comparable to the binding energy of the CGM, then the CGM will expand, and its pressure will decrease.  A feedback mechanism with these properties tunes the pressure of the galaxy's CGM to keep the electron density profile within the galaxy close to $n_{e,{\rm eq}}$.  
\vspace*{1em}
\label{schematic_nel_plot_240_models}}
\end{figure*}

The general proximity of these heating/cooling ratio profiles to unity is remarkable.  Naively, one might expect gas below the line of equality to cool further, becoming increasingly compressed and cooling-dominated.  Likewise, heating of gas above the line of equality should cause it to expand and become more strongly dominated by heating.  The nearness of these observed profiles to the line of heating/cooling equality therefore indicates that a self-regulation mechanism involving AGN feedback maintains these galaxies near the balance point between stellar heating and radiative cooling.  Also notable is the fact that the heating/cooling ratio profiles beyond $\sim 1$~kpc for galaxies without extended multiphase gas (left panel) tend to rise from $r \sim 1$~kpc to $r \sim 10$~kpc, while the profiles of ellipticals that do contain extended multiphase gas (right panel) tend either to decline or to remain flat.

\subsection{Boundary Pressure and Cooling Flows}
\label{sec-BoundaryPressure}

In steady subsonic solutions for thermally driven, mass-loaded outflows, the normalization of pressure and density depend on the gas pressure at the outer boundary.  For simplicity, let $r_{\rm b}$ be a boundary radius inside of which stellar heating determines the outflow's entropy profile.\footnote{There is no clear physical distinction between the interstellar medium (ISM) of a massive elliptical galaxy and its CGM, and so this boundary radius is rather arbitrary and imprecise.  We expect it to be similar to the effective radius $r_{\rm eff}$ enclosing half of the galaxy's stellar mass.  In practice, however, we infer it from the morphology of the galaxy's entropy profile (see \S \ref{sec-ObsComparison}).}  Setting the bounding gas pressure at $r_{\rm b}$ to $P_{\rm b}$ then leads to the profiles 
\begin{eqnarray}
  P_{\rm out}(r) & \, \approx \, & P_{\rm b} \left( \frac {r} {r_{\rm b}} \right)^{- 3 \alpha_K / 2} 
  	\label{eq-P_out} \\
  kT_{\rm out}(r) & \, \approx \, & kT_{\rm b} 
  	\label{eq-kT_out} \\
  n_{e,{\rm out}}(r) & \, \approx \, &  \frac {P_{\rm b}} {kT_b}  \left( \frac {\mu} {\mu_e} \right) 
  			 \left( \frac {r} {r_{\rm b}} \right)^{- 3 \alpha_K / 2}  
	\label{eq-ne_out} \\
  K_{\rm out}(r) & \, \approx \, & \left( k T_b \right)^{5/3} 
                          \left( \frac {P_{\rm b}} {\mu_e / \mu} \right)^{-2/3} 
  			 \left( \frac {r} {r_{\rm b}} \right)^{\alpha_K} 
  	\label{eq-K_out} 
	  \; \; ,
\end{eqnarray}
where $\alpha_K$ is determined by equation (\ref{eq-alpha_K_vc}), $\mu_e m_p$ is the mean mass per electron, and $kT_{\rm b} = 2 \mu m_p v_c^2 / 3 \alpha_K$.

In galaxies with $\sigma_v \approx 240 \, {\rm km \, s^{-1}}$, these supernova-sweeping profiles run parallel to the critical profiles for heating-cooling balance.  Whether or not the region inside of $r_{\rm b}$ lies either entirely above or entirely below the critical profile $P_{\rm eq} (r)$ depends directly on the boundary pressure (see Figure \ref{schematic_nel_plot_240_models}).  The AGN feedback in a galaxy like this can shut off a galaxy-wide cooling flow simply by heating and lifting its CGM until $P(r_{\rm b}) < P_{\rm eq}$.  After that happens, the shutdown of the cooling flow may not be permanent, because ejected stellar mass swept out of the galaxy continues to accumulate in the CGM.  Unless a nonstellar energy source continually heats, strips, or lifts the accumulating gas, the CGM pressure will gradually increase until $P(r_{\rm b}) > P_{\rm eq}$, at which time the resulting cooling flow needs to trigger another feedback episode that lowers $P_{\rm b}$.

\begin{figure*}[t]
\begin{center}
\includegraphics[width=7.0in, trim = 0.0in 0.2in 0.0in 0.0in]{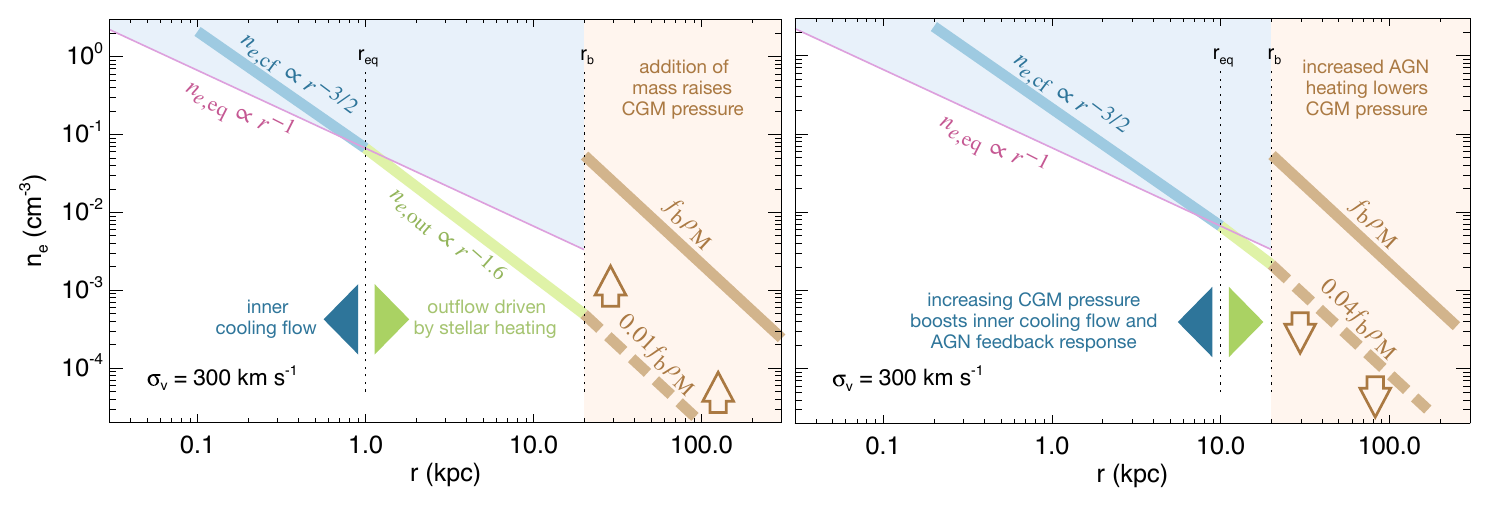} \\
\end{center}
\caption{ \footnotesize 
Schematic illustration of the black hole feedback valve in a galaxy with $\sigma_v \approx 300 \, {\rm km \, s^{-1}}$.  All figure elements in common with Figure \ref{schematic_nel_plot_240_models} have the same meanings.  However, the density profile of this galaxy's outflow ($n_{e,{\rm out}}$) intersects the locus of heating/cooling equality ($n_{e,{\rm eq}}$) at a radius $r_{\rm eq}$ determined by the boundary pressure.  In the left panel, the boundary pressure is 1\% of the cosmological expectation, causing $r_{\rm eq}$ to be at 1~kpc. Inside of that radius, a cooling flow carries ejected stellar gas inward (blue arrowhead).  Outside of that radius, stellar heating causes an outflow that transports ejected stellar gas into the CGM (green arrowhead).  In the right panel, the CGM pressure has increased to 4\% of the cosmological expectation, and the radius of equality has moved outward to 10~kpc, causing the inward mass flow rate to increase and to fuel stronger AGN feedback.  One expects AGN feedback in such a galaxy to adjust the CGM pressure so that expansion of the CGM by thermalization of feedback energy offsets the increases in CGM mass that would otherwise come from cosmological accretion and outflows of ejected stellar gas.
\vspace*{1em}
\label{schematic_nel_plot_300_models}}
\end{figure*}

\subsection{The Valve}
\label{sec-TheValve}

The mechanism depicted in Figure \ref{schematic_nel_plot_240_models} closely links the strength of AGN feedback with the confining CGM pressure.  It is essentially a switch that turns on a central cooling flow when the CGM pressure is high and turns the cooling flow off when the CGM pressure is low.   The onset of a cooling flow presumably triggers AGN feedback, but the feedback response does not necessarily need to limit cooling by heating gas at small radii.  Instead, it can limit cooling at small radii by lowering the CGM pressure, which allows the ambient gas at 1--10~kpc to expand, temporarily making local stellar heating more competitive with local radiative cooling.  The AGN feedback in a galaxy with this feature will tend to drive the boundary pressure toward the locus of heating/cooling equality, so that $P_{\rm b} \sim P_{\rm eq} (r_{\rm b})$.

In galaxies with $\sigma_v \gtrsim 240 \, {\rm km \, s^{-1}}$ this regulation mechanism becomes more like an adjustable valve than a switch.  Figure~\ref{schematic_nel_plot_300_models} shows how the valve works.  An outflow driven by stellar heating in a galaxy with this larger velocity dispersion has a density profile ($n_{e,{\rm out}}$) that is steeper than the locus of heating/cooling equality ($n_{e,{\rm eq}}$).  Those profiles intersect at a particular radius ($r_{\rm eq}$).  Inside of that radius is a cooling flow, and outside of it is an outflow driven by stellar heating.  The mass inflow rate of the cooling flow ($\dot{M}_{\rm cf}$) is equivalent to the combined stellar mass-loss rate of all the stars within $r_{\rm eq}$, which is determined by the confining CGM pressure.  Therefore, the time-averaged rate at which the galaxy supplies gas to its central black hole is a continuous function of the CGM pressure, with greater CGM pressure resulting in a greater fueling rate.  The outcome is a mechanism capable of tuning itself so that kinetic AGN feedback regulates the confining CGM pressure to remain near a time-averaged level consistent with the required AGN fueling rate.

The mechanism requires kinetic energy output from the AGN to offset three different effects that act to increase CGM pressure.
\begin{enumerate}

\item \textit{CGM Cooling.}  Radiative losses from the CGM around isolated galaxies with $\sigma_v \gtrsim 240 \, {\rm km \, s^{-1}}$ are typically $10^{41} \, {\rm erg \, s^{-1}} \lesssim L_{\rm X} \lesssim 10^{42} \, {\rm erg \, s^{-1}}$.  The time-averaged energy supply from the AGN must be at least this large in order to prevent CGM pressure from increasing because of a gradual decline in entropy and a gradual increase in density. 

\item \textit{Stellar Gas Ejection.}  As mentioned in \S 2.5, supernova heating alone cannot unbind ejected stellar gas from the dark matter halo around a galaxy with $\sigma_v \gtrsim 240 \, {\rm km \, s^{-1}}$.  Continual lifting to large altitude therefore requires an additional power input of $\sim 4.6 M_* v_c^2 / t_*$ from the AGN,\footnote{The factor of 4.6 approximates the work required to lift gas in an isothermal potential to $\sim 100$ times its original altitude, or equivalently, the work required to unbind gas originally at the bottom of an NFW potential well.} corresponding to $\sim 10^{41.5} \, {\rm erg \, s^{-1}} \, (M_* / 10^{11.3} \, M_\odot) \, \sigma_{240}^2$.  

\item \textit{Cosmological Gas Accretion.}  The time-averaged cosmological infall rate of gas into the CGM is $\sim f_b M_{\rm halo} H_0$, which substantially exceeds the mass input from stars.  Lifting of that CGM gas within the halo's gravitational potential\footnote{Heating that adds a specific energy $\sigma_v^2$ to CGM gas in an isothermal potential without changing the slope of its entropy profile lifts nearly hydrostatic gas by a factor of $e^{1/2} \approx 1.6$ in radius and lowers its density by a factor of $\sim 5$.} requires the AGN to supply a time-averaged power of $\sim 10^{43} \, {\rm erg \, s^{-1}} \, (M_{\rm halo} / 10^{13} \, M_\odot) \, \sigma_{240}^2$.

\end{enumerate}
The most demanding job for AGN feedback in this halo mass range is therefore to lift the CGM gas that accumulates through cosmological accretion.

\begin{figure*}[t]
\begin{center}
\includegraphics[width=7.0in, trim = 0.0in 0.2in 0.0in 0.0in]{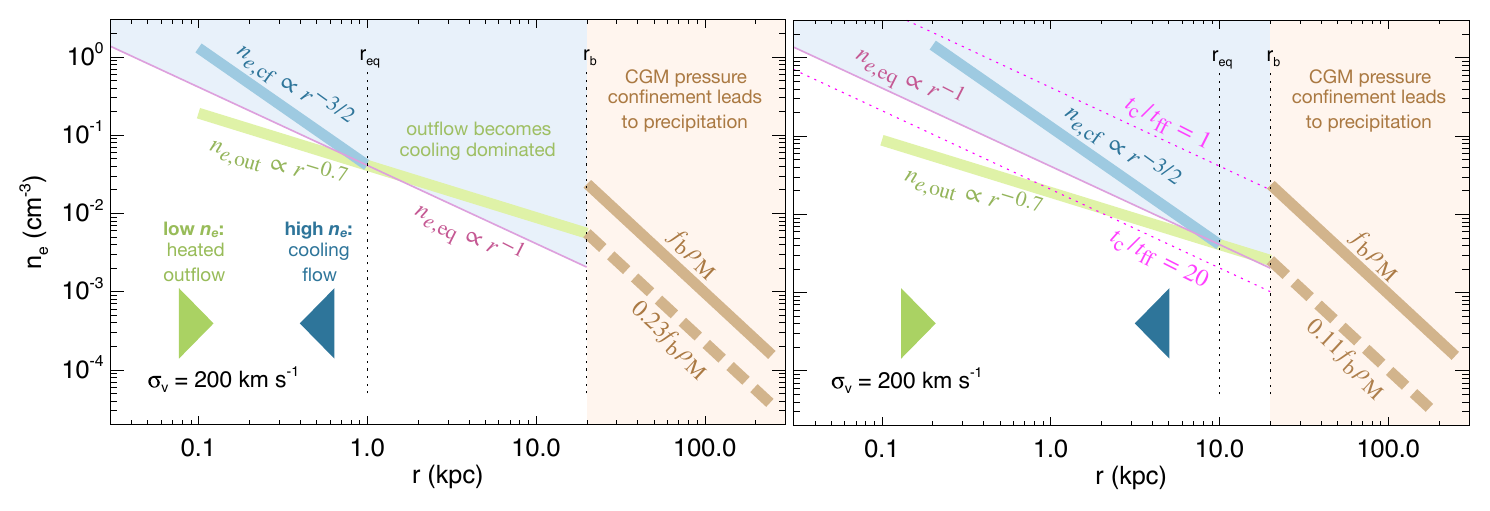} \\
\end{center}
\caption{ \footnotesize 
Schematic illustration showing why supernova-heated outflows in galaxies with $\sigma_v \approx 200 \, {\rm km \, s^{-1}}$ are susceptible to multiphase circulation.  All figure elements in common with Figure~\ref{schematic_nel_plot_300_models} have the same meanings.  The boundary pressure no longer uniquely determines the flow pattern at smaller radii because the slope of $n_{e,{\rm out}}$ is shallower than the slope of $n_{e,{\rm eq}}$. Both steady outflow and steady inflow solutions are possible within $r_{\rm eq}$, as indicated by placement of both blue and green arrowheads at smaller radii than $r_{\rm eq}$.   However, neither flow can remain homogeneous.  Stellar heating can power an outflow that begins with $n_e < n_{e,{\rm eq}}$ at small radii (green line), but the density profile of outflowing gas eventually intersects $n_{e,{\rm eq}}$ at $r_{\rm eq}$.  Beyond that point, radiative cooling will cause the entropy of the outflow to drop, producing an entropy inversion that is unstable to convection.   Multiphase condensation should therefore occur near $r_{\rm eq}$, causing cold clouds to rain inward through the hot outflow.  On the other branch (blue line), the $t_{\rm cool} / t_{\rm ff}$ ratio of a steady cooling flow declines as the gas moves inward, eventually falling to a point at which multiphase condensation ensues.  For reference, the upper dotted magenta line in the right panel indicates $t_{\rm cool} / t_{\rm ff} = 1$ for $kT = \mu m_p v_c^2$ and the lower one indicates $t_{\rm cool} / t_{\rm ff} = 20$ at the same temperature.
\vspace*{1em}
\label{schematic_nel_plot_200_models}}
\end{figure*}

Integrated over cosmic time, the total amount of energy required to alleviate the galaxy's boundary pressure by continually lifting accreting CGM gas is $\sim f_{\rm b} M_{\rm halo} \sigma_v^2$, equal to $\sim 10^{60.3} \, {\rm erg} \, (M_{\rm halo} / 10^{13} \, M_\odot) \sigma_{240}^2$.  During the same time period, the central black hole grows to have a rest-mass energy $\sim (10^{63} \, {\rm erg}) \, \sigma_{240}^{4.4}$, based on observations indicating that $M_{\rm BH} \approx (7 \times 10^8 \, M_\odot) \, \sigma_{240}^{4.4}$ \citep{KormendyHo2013ARAA..51..511K}.  Tapping less than 1\% of the central black hole's rest-mass energy as it grows therefore suffices to lift the CGM (see also \S \ref{sec-MBH_sigmav}).

\subsection{Multiphase Circulation}
\label{sec-ThermalStability}

Star formation in a galaxy resembling the left panel of Figure~\ref{schematic_nel_plot_300_models} is effectively quenched because the outflow that carries away stellar ejecta remains homogeneous and the inner cooling flow produces less than $1 \, M_\odot \, {\rm yr}^{-1}$ of condensed gas.  However, outflows from galaxies with $\sigma_v \lesssim 240 \, {\rm km \, s^{-1}}$ are unlikely to be stable to multiphase condensation.  Figure~\ref{schematic_nel_plot_200_models} illustrates the problem.  A heated outflow of stellar ejecta through a shallow potential well tends to have a density slope that is shallower than $n_{e,{\rm eq}}$ and therefore passes from a locally heating-dominated state to a locally cooling-dominated state at $r_{\rm eq}$.  An entropy inversion then develops outside of $r_{\rm eq}$ on a timescale of $\sim t_{\rm cool} (r_{\rm eq})$.  That inversion is unstable to convection, which allows low-entropy gas to condense out of the outflow and to rain back toward the galaxy's center.  Multiphase circulation therefore develops and can resupply the galaxy with cold clouds capable of forming stars.  

More generally, the ability of a galactic atmosphere to remain homogeneous depends heavily on its ambient $t_{\rm cool}/t_{\rm ff}$ ratio 
\citep[e.g.,][]{Hoyle_1953ApJ...118..513H,Nulsen_1986MNRAS.221..377N,bs89,bnf09,McCourt+2012MNRAS.419.3319M,Sharma_2012MNRAS.420.3174S,
Thompson_2016MNRAS.455.1830T,Voit_2017_BigPaper,ChoudhurySharma_2016MNRAS.457.2554C,Choudhury_2019MNRAS.488.3195C}.  In the conventional definition of this ratio, the cooling time is $t_{\rm cool} \equiv  3 P / 2 n_e n_p \Lambda (T)$, where $\Lambda(T)$ is the usual radiative cooling function, and the freefall time is $t_{\rm ff} \equiv (2 r / g)^{1/2}$, where $g$ is the local gravitational acceleration.  A static, thermally balanced medium can, in principle, be stable to multiphase condensation if $t_{\rm cool} / t_{\rm ff} \gtrsim 1$, because buoyancy effects enabled by entropy stratification suppress condensation \citep{Cowie_1980MNRAS.191..399C,Nulsen_1986MNRAS.221..377N,bnf09,McCourt+2012MNRAS.419.3319M}.  In practice, however, media with $1 \lesssim t_{\rm cool} / t_{\rm ff} \lesssim 10$ remain highly susceptible to multiphase condensation because subsonic disturbances can interfere with the condensation-damping effects of buoyancy
\citep{Sharma_2012MNRAS.420.3174S,Gaspari+2013MNRAS.432.3401G,Voit_2017_BigPaper}, particularly if those disturbances flatten or invert the entropy gradients that are responsible for buoyancy \citep{McNamara_2016ApJ...830...79M,Voit_2017_BigPaper,Choudhury_2019MNRAS.488.3195C}.   

The result is a pervasive upper limit on the pressure and density of ambient gas corresponding to $\min(t_{\rm cool}/t_{\rm ff}) \approx 10$, which we will call the precipitation limit
\citep{Sharma+2012MNRAS.427.1219S,Voit_2015Natur.519..203V,Voit2018_LX-T-R,Voit_2019ApJ...879L...1V,Hogan_2017_tctff,Babyk_2018ApJ...862...39B,Voit_2019ApJ...880..139V}.   Galactic atmospheres tend to have $t_{\rm cool} / t_{\rm ff} \gtrsim 10$ because ambient gas with a shorter cooling time cannot persist indefinitely without forming condensates that fuel feedback.  As the cooling time rises, so that $t_{\rm cool} / t_{\rm ff}$ grows from $\sim 10$ to $\sim 20$, increasingly large hydrodynamical disruptions are required to offset the buoyancy effects that suppress multiphase condensation \citep{Voit_2017_BigPaper,Voit_2018ApJ...868..102V}.  And if $t_{\rm cool} / t_{\rm ff} \gtrsim 20$, then multiphase condensation cannot occur without either large entropy perturbations \citep{Choudhury_2019MNRAS.488.3195C}, strong magnetic fields \citep{Ji_2018MNRAS.476..852J}, or enough rotational support to offset the stabilizing effects of buoyancy \citep{Gaspari_2015A&A...579A..62G,SobacchiSormani_2019MNRAS.486..205S,SormaniSobacchi_2019MNRAS.486..215S}.  A subsonic outflow is therefore likely to remain stable to multiphase condensation if it has (1) an entropy gradient with $\alpha_K \gtrsim 2/3$, (2) $t_{\rm cool} / t_{\rm ff} \gtrsim 20$ at all radii, and (3) a lack of large entropy perturbations, strong magnetic fields, or significant rotational support. 

The right panel of Figure~\ref{schematic_nel_plot_200_models} shows how $n_{e,{\rm eq}}$ relates to $t_{\rm cool} / t_{\rm ff}$ in a galaxy with $\sigma_v \approx 200 \, {\rm km \, s^{-1}}$.  A green line shows the outflow solution ($n_{e, {\rm out}}$) obtained by requiring heating to equal cooling at 10~kpc.  As the outflow approaches that radius, it is already at $t_{\rm cool} / t_{\rm ff} < 20$ and is approaching $t_{\rm cool} / t_{\rm ff} \approx 10$.  One therefore expects the outflow to become unstable to multiphase condensation in the neighborhood of $r_{\rm eq}$.  In other words, star formation in lower-mass galaxies is more difficult to quench with feedback, because the outflows that feedback generates tend to destabilize the CGM.  Similarly, a cooling flow starting at $r_{\rm eq}$ becomes increasingly susceptible to multiphase condensation as it moves inward because its density slope ($n_{e,{\rm cf}} \propto r^{-1.5}$) is steeper than the lines indicating constant $t_{\rm cool} / t_{\rm ff}$.  The ambient $t_{\rm cool} / t_{\rm ff}$ ratio therefore decreases as the flow moves inward, and the flow cannot remain homogeneous within the radius at which $t_{\rm cool} \approx t_{\rm ff}$.

\subsection{Jet Propagation}
\label{sec-JetPropagation}

In order for the black hole feedback valve to operate as described, most of the AGN feedback energy produced over cosmic time needs to be thermalized at radii larger than where most of the galaxy's stars reside.  Morphological observations of strong radio jets indicate that they can indeed propagate as narrow streams beyond most of the stars.  For example, NGC 4261 (represented by purple stars in Figure 1) has radio jets that are narrow out to $r > 10$~kpc and terminate in lobes at $r \sim 25$--45~kpc, consistent with a power output $> 10^{43} \, {\rm erg \, s^{-1}}$\citep{OSullivan_2011MNRAS.416.2916O}.  Meanwhile, that galaxy's profiles of gas density, temperature, and entropy at $r < 10$~kpc remain nearly identical to those of the other galaxies with $\sigma_v \sim 300 \, {\rm km \, s^{-1}}$, even though its jets are unusually powerful.

Propagation to such a distance requires the momentum flux in the jets to be at least as great as the ambient pressure, which is $\sim 10^{-11} \, {\rm erg \, cm^{-3}}$ at $\sim 10$~kpc in NGC 4261 \citep{OSullivan_2011MNRAS.416.2916O}.  For comparison, the relativistic momentum flux corresponding to a jet power $\dot{E}_{42} \times 10^{42} \, {\rm erg \, s^{-1}}$ is
\begin{equation}
  \frac {3 \times 10^{-12} \, {\rm erg \, cm^{-3}}} {\Omega_{\rm jets}}
     \left( \frac {\dot{E}_{42}} {100} \right)
    \left( \frac {r} {10 \, {\rm kpc}} \right)^{-2}
	\label{eq-MomFlux}
     \; \; ,
\end{equation}
where $\Omega_{\rm jets}$ is the solid angle in steradians that the jets subtend at radius $r$.  Observations indicate that the jet opening angle at $\sim 10$~kpc in NGC 4261 is $\sim 12^\circ$\citep{Nakahara_2018ApJ...854..148N}, implying that the combined solid angle of both jets is $\sim 0.3$~sr and that a jet power $\gtrsim 10^{44} \, {\rm erg \, s^{-1}}$ is needed to drill through the gas at $r < 10$~kpc.  

Jet power in the other galaxies shown in Figure \ref{Werner_equality_red_blue} is far smaller, typically $\sim 10^{41}$--$10^{42} \, {\rm erg \, s^{-1}}$ \citep{Werner+2014MNRAS.439.2291W}.  Equation (\ref{eq-MomFlux}) implies that those jets are currently incapable of propagating at relativistic speeds to distances $\gtrsim 10$~kpc unless they are extremely narrow, and their observed morphologies indicate that they do not extend to $\gtrsim 10$~kpc as narrowly collimated outflows.  Instead, the jets thermalize their kinetic energy and inflate bubbles at smaller radii as they decelerate, sometimes driving shocks that propagate into the supernova-heated outflow.   According to V15, those shocks should be relatively weak, each imparting an entropy jump
\begin{equation}
 \Delta K_{\rm jets} \sim (2.8 \, {\rm keV \, cm^2})  \, 
 		\dot{E}_{42}^{2/3}
 		\Lambda_ {\rm 3e-23}^{4/3} \, 
		\sigma_{240}^{-4}
		\label{eq-K_jets}
		\; \; ,
\end{equation}
with $\Lambda_{\rm 3e-23} \equiv \Lambda / (3 \times 10^{-23} \, {\rm erg \, cm^3 \, s^{-1}})$, as they propagate through an atmosphere near the precipitation limit.  Shocks driven by $\sim 10^{42} \, {\rm erg \, s^{-1}}$ of power become subsonic at $\sim 10$~kpc.  Beyond that radius, the hot bubbles that drove the shocks are expected to buoyantly rise and eventually to mix with the ambient medium, thermalizing much of the original AGN power at greater altitudes \citep[e.g.,][]{Churazov+01,Birzan+04,vd05}.   As a result, AGN feedback in massive elliptical galaxies does not substantially alter the entropy profile slope established within $\lesssim 10$~kpc by a supernova-heated outflow, except at $\lesssim 1$~kpc, where the observed core entropy is comparable to $\Delta K_{\rm jets}$ (see V15 and \S \ref{sec-ObsComparison}).

\section{Steady Flow Solutions}
\label{sec-SteadyFlow}

Section~\ref{sec-BasicPicture} considered two energy sources with the potential to offset radiative cooling and limit star formation.  Heating by SN Ia is relatively steady and has a known spatial distribution but is not energetically capable of pushing ejected stellar gas all the way out of the galaxy's potential well.  On the other hand, AGN feedback can provide enough energy to push away all of the gas associated with a massive galaxy but is intermittent in time, with a poorly known spatial distribution.  This section therefore considers the steady flow patterns produced by SN Ia heating alone.  Section~\ref{sec-ObsComparison} then follows up by attempting to infer the contributions of AGN feedback from the observed discrepancies between the steady flow solutions and observations of the atmospheres of real galaxies.

The fluid equations for steady one-dimensional radial flow are
\begin{eqnarray}
    \frac {1} {r^2} \frac {d} {dr} (\rho v_r r^2 ) & \; = \; & \dot{\rho}(r)      \label{eq-mass}  \\
    \frac {1} {r^2} \frac {d} {dr} (\rho v_r^2 r^2 ) & \; = \; & - \frac {dP} {d r} - \rho \frac {d\phi} {dr}   \label{eq-momentum}  \\
    \frac {1} {r^2} \frac {d} {dr} (\epsilon \rho v_r r^2 )  & \; = \; & {\cal H}(r) + \dot{\rho}(r) \phi (r)      \label{eq-energy}  
    \, \, .
\end{eqnarray}
In a galactic environment, $\dot{\rho}(r) = \rho_*(r) / t_*$ is a source term for stellar mass loss, and ${\cal H}(r)$ is the net stellar heating rate per unit volume.  The standard integration method for steady galactic flows is to choose a stagnation radius ($r_0$) and integrate away from it in both directions, iteratively seeking a solution that passes smoothly through a cooling-flow sonic point at small radii \citep[e.g.,][]{David_Part1_1987ApJ...313..556D,TaborBinney1993MNRAS.263..323T}.   Physically, the steady-state flow configuration and its stagnation radius are determined by the CGM boundary pressure, along with an implicit assumption that some mechanism other than stellar heating keeps the pressure at $r_{\rm b}$ constant by removing the gas that flows to greater radii.  

Figure~\ref{schematic_K_plot_models} illustrates how steady flow solutions without AGN heating\footnote{Adding an amount of AGN heating comparable to stellar heating and with a similar spatial distribution would tend to decrease the outflow's entropy slope $\alpha_K$.}  depend on a galaxy's central stellar velocity dispersion $\sigma_v$.  A teal line in each panel shows the entropy profile corresponding to a steady flow solution determined by only two parameters: $\sigma_v$ and either $r_0$ or the boundary pressure $P_{\rm b}(r_{\rm b})$.  The figure shows steady flow solutions with $t_{\rm cool} \sim 1 \, {\rm Gyr}$ at 10~kpc because X-ray observations of massive elliptical galaxies typically indicate $t_{\rm cool} \sim 0.5$ to $\sim 2$~Gyr at 10~kpc \citep{Lakhchaura_2018MNRAS.481.4472L}.  In the four panels representing galaxies with $\sigma_v \geq 230 \, {\rm km \, s^{-1}}$, the steady solution shown consists of a cooling flow inside of $r_0$ and a supernova-heated outflow outside of $r_0$.  A kink near $r_{\rm eq}$ indicates the location of the stagnant region, and the local entropy minimum in that region is at $r_0$.  Inside of $r_0$, the steady flow solutions in those four panels resemble pure cooling flows, since they have $K \propto r$, as appropriate for a steady cooling flow in an isothermal potential.  At larger radii ($r > r_0$), the steady flow solutions resemble supernova-heated outflow solutions, with $\alpha_K$ close to the prediction from equation (\ref{eq-alpha_K_vc}), meaning that the outflow's entropy slope increases as $\sigma_v$ increases.  

\begin{figure*}[t]
\begin{center}
\includegraphics[width=6.0in, trim = 0.0in 0.0in 0.0in 0.0in]{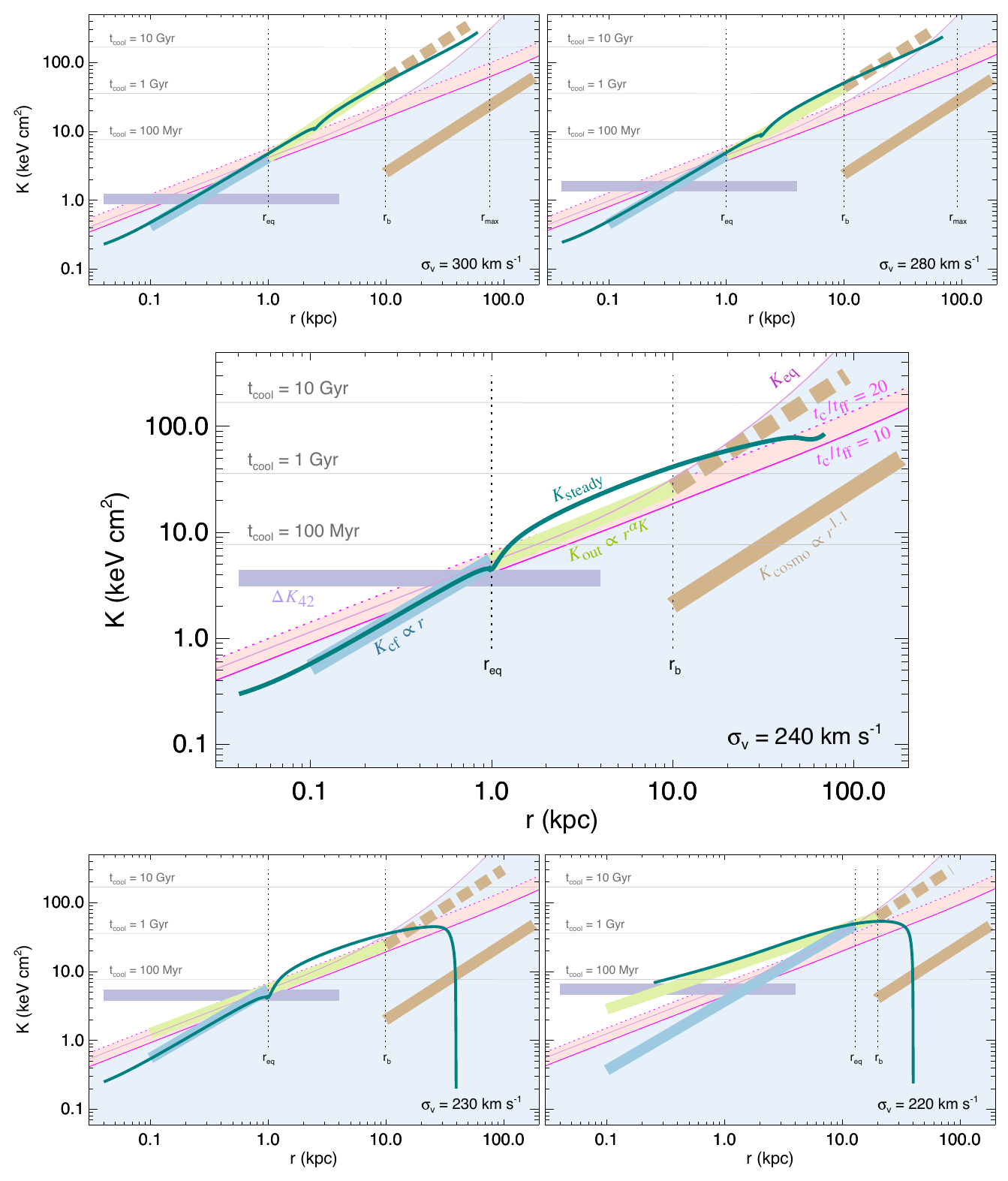} \\
\end{center}
\caption{ \footnotesize 
Entropy profiles of representative steady flow solutions for generic massive galaxies.  In each panel, a teal line (labeled $K_{\rm steady}$ in the central panel) shows a particular steady flow solution for a galaxy with the central stellar velocity dispersion ($\sigma_v$) given in the label.  A thick blue line ($K_{\rm cf}$) shows the entropy slope expected for a pure cooling flow.  A thick green line ($K_{\rm out}$) shows the entropy slope expected for an outflow driven by SN Ia heating.  A thick solid tan line ($K_{\rm cosmo}$) shows the entropy profile produced around the galaxy by pure cosmological structure formation.   A thick dashed tan line shows the same profile displaced upward in entropy so that it is continuous with $K_{\rm out}$.   A thin violet line shows the entropy level $K_{\rm eq}$ at which radiative cooling would balance SN Ia heating.  Cooling exceeds SN Ia heating in the blue region below the violet line.  A thin solid magenta line shows the precipitation limit at $t_{\rm cool} / t_{\rm ff} \approx 10$.  A thin dotted magenta line shows $t_{\rm cool} / t_{\rm ff} \approx 20$.  The pink region between those lines corresponds to intermediate values of $t_{\rm cool} / t_{\rm ff}$ that may be susceptible to multiphase condensation.  Thin horizontal gray lines indicate where $t_{\rm cool} = 100$~Myr, 1~Gyr, and 10~Gyr, as labeled.  A thick purple line from 0.04 to 4~kpc in each panel shows the entropy jump ($\Delta K_{42}$) that a spherical shock driven by a $10^{42} \, {\rm erg \, s^{-1}}$ outflow would produce in a medium at the precipitation limit.  Vertical black dotted lines in each of the upper two panels show the maximum radius $r_{\rm max}$ to which stellar heating can drive ejected stellar gas.  
\vspace*{1em}
\label{schematic_K_plot_models}}
\end{figure*}

In galaxies with $\sigma_v \lesssim 240 \, {\rm km \, s^{-1}}$, the steady outflow solutions develop unsustainable entropy inversions after passing from the heating-dominated region of the $r$--$K$ plane into the cooling-dominated region.  As expected from the analysis of \S 2, the outflow solutions for those galaxies have entropy slopes too shallow to remain in the heating-dominated region.  After the outflow solution enters the cooling dominated region, its entropy profile starts to decline, implying that the flow has become unstable to convection.  As the entropy continues to decline with radius, the flow encounters the solid magenta line marking $t_{\rm cool} / t_{\rm ff} \approx 10$.  Near that point, the flow should become unstable to multiphase condensation, enabling low-entropy gas blobs to precipitate out of the flow and to start sinking inward.  In galaxies with these properties, outflows driven by stellar heating therefore drive multiphase circulation instead of homogeneously expelling the gas being shed by stars.

The remainder of this section interprets the solutions illustrated in Figure~\ref{schematic_K_plot_models} in more detail.  First, it outlines the galaxy model used to specify the gravitational potential and the source terms for mass and energy input, which depend only on $\sigma_v$.  Then, it discusses each panel of the figure.  Readers more interested in comparisons with observations may wish to skip to \S \ref{sec-ObsComparison}.

\subsection{Generic Galaxy Model}
\label{sec-GenericGalaxy}

The generic galaxy model used in the steady flow integrations assumes that the galaxy's halo has a Navarro-Frenk-White NFW density profile, $\rho_{\rm M} \propto (r/r_{\rm s})^{-1} (1 + r/r_{\rm s})^{-2}$, with a maximum circular velocity $v_{c,{\rm max}} = \sqrt{2} \,  \sigma_v$ and a scale radius $r_s = 0.1 r_{200}$, where $r_{200}$ is the radius encompassing a mean mass density 200 times the cosmological critical density.  It also assumes that the distribution of stellar mass density follows a modified Einasto profile with 
\begin{equation}
  \frac {d \ln \rho_*} {d \ln r} =  - 2 \left( \frac {r} {r_{-2}} \right)^{1/n_{\rm E}}
  \label{eq-ModEinasto}
 \end{equation}
for $r \geq r_{-2}$ and $\rho_* (r) \propto r^{-2}$ for $r < r_{-2}$.  The index describing the outer stellar envelope is set to $n_{\rm E} = 4$, the stellar mass density profile is normalized so that $\rho_*(r_{-2}) = \sigma_v^2 / 2 \pi G r^2$, and its scale radius $r_{-2}$ is determined by setting the two-dimensional effective radius of the galaxy equal to $0.015 r_{200}$, in approximate agreement with the mean relation observed among galaxies \citep{Kravtsov_2013ApJ...764L..31K}.  This prescription results in an overall potential well with a circular velocity that is nearly constant with radius, remaining within 10\% of $v_{\rm c} = \sqrt{2} \,  \sigma_v$ out to $0.75 \, r_{200}$, in alignment with the assumptions of \S \ref{sec-BasicPicture}.  However, comparable galaxies at the centers of higher-mass groups and clusters sit in potentials with greater maximum circular velocity.  Alleviating the boundary pressure around those galaxies by lifting the CGM therefore requires substantially more AGN feedback energy (see \S \ref{sec-ClusterCores}).

\subsection{Description of Solutions}

\subsubsection{$\sigma_v = 300 \, {\rm km \, s^{-1}}$}

The most massive generic galaxy shown in Figure~\ref{schematic_K_plot_models} has a velocity dispersion $\sigma_v = 300 \, {\rm km \, s^{-1}}$.  Its total stellar mass is $10^{11.8} \, M_\odot$, and its total mass within $r_{200}$ is $M_{200} \approx 10^{13.2} \, M_\odot$.  In this potential well, stellar heating alone cannot push ejected stellar gas beyond $r_{\rm max} \approx 70$~kpc.  The steady flow solution depicted for this galaxy is determined by the boundary condition $r_0 = 2.5$~kpc, so that $r_{\rm eq} \approx 1$~kpc, in alignment with Figure~\ref{Werner_equality_red_blue}.  Gas between $r_{\rm eq}$ and $r_0$ is experiencing net heating but is flowing inward because it rests upon a cooling flow that is moving inward.  The cooling-flow rate reaches $\dot{M}_{\rm cf} \approx 0.5 \, M_\odot \, {\rm yr}^{-1}$ at small radii because the stellar mass within $r_0$ is $10^{11} \, M_\odot$ and the specific stellar mass-loss rate is $(200 \, {\rm Gyr})^{-1}$.  The figure shows that the entropy slope of the inner cooling flow is nearly identical to the $K \propto r$ expectation (thick blue line) for a cooling flow in an isothermal potential.  

Outside of $r_0$, equation (\ref{eq-alpha_K_vc}) predicts an entropy slope $\alpha_K = 1.1$, shown by the thick green line, and the steady outflow solution there has a similar slope.  It therefore climbs progressively higher into the heating-dominated region as it moves outward, reaching a cooling time $\sim 2$~Gyr at $r_{\rm b} = 10$~kpc, where its pressure is $P_{\rm b} \approx 3 \times 10^{-12} \, {\rm erg \, cm^{-3}}$.  Beyond $r_{\rm b}$, the thick dashed tan line shows a continuation of $K(r)$ with a cosmological entropy slope $K \propto r^{1.1}$ and a normalization $\sim 20$ times the entropy profile expected from pure cosmological structure formation (thick solid tan line).  This steady flow solution therefore requires AGN feedback or some other mechanism to reduce the CGM pressure at $r_{\rm b}$ by a factor of $\sim 10^{-2}$ (because $P \propto K^{-3/2}$ at a given $T$).

\subsubsection{$\sigma_v = 280 \, {\rm km \, s^{-1}}$}

The second most massive galaxy shown in Figure~\ref{schematic_K_plot_models} has a velocity dispersion $\sigma_v = 280 \, {\rm km \, s^{-1}}$, a total stellar mass of $10^{11.7} \, M_\odot$, and $M_{200} \approx 10^{13.1} \, M_\odot$.  A boundary condition $r_0 = 2$~kpc determines the steady flow solution in the figure.  Stellar heating cannot push ejected stellar gas beyond $r_{\rm max} \approx 90$~kpc, and the inner cooling-flow rate reaches $\dot{M}_{\rm cf} \approx 0.4 \, M_\odot \, {\rm yr}^{-1}$. 

Again, the entropy slope of the outflow beyond $r_0$ is similar to the prediction of equation (\ref{eq-alpha_K_vc}), which in this case is $\alpha_K = 0.9$.  This slope is steep enough to remain in the heating-dominated region of the $r$--$K$ plane out to tens of kpc.   The flow's cooling time increases through $\sim 2$~Gyr at $r_{\rm b} = 10$~kpc, where its pressure is $P_{\rm b} \approx 3 \times 10^{-12} \, {\rm erg \, cm^{-3}}$.  As in the $\sigma_v = 300 \, {\rm km \, s^{-1}}$ case, the CGM pressure at $r_{\rm b}$ must be reduced by a factor of $\sim 10^{-2}$ below its cosmological value in order for the solution to be valid.

\subsubsection{$\sigma_v = 240 \, {\rm km \, s^{-1}}$}

The galaxy with a critical velocity dispersion ($\sigma_v = 240 \, {\rm km \, s^{-1}}$) has a total stellar mass of $10^{11.53} \, M_\odot$ and $M_{200} \approx 10^{12.89} \, M_\odot$.  Setting $r_0 = 1.5$~kpc determines the steady flow solution shown.  Stellar heating cannot push ejected stellar gas beyond $r_{\rm max} \approx 160$~kpc, and the inner cooling-flow rate reaches $\dot{M}_{\rm cf} \approx 0.2 \, M_\odot \, {\rm yr}^{-1}$.   Just beyond $r_0$, the unique requirements of a steady-state solution cause the entropy profile to climb steeply out to $\sim 2$~kpc, but the outflow solution from 2 to 10 kpc has a slope similar to the prediction $\alpha_K = 2/3$ from equation (\ref{eq-alpha_K_vc}).  In order for this solution to be valid, the CGM pressure at $r_{\rm b}$ must be reduced by a factor of $\sim 10^{-2}$ below its cosmological value. 

Beyond 10~kpc, something interesting happens.  The steady outflow solution crosses back into the cooling-dominated region of the $r$--$K$ plane.  Its entropy slope therefore becomes progressively shallower until it starts to decline.  The resulting entropy inversion is unstable to convection, meaning that a steady homogeneous outflow cannot be sustained.  Lower-entropy gas at larger radii cannot be hydrostatically supported and becomes unstable to multiphase condensation.  Multiphase circulation near $r_{\rm b}$ consisting of low-entropy gas clouds descending through a higher-entropy outflow is therefore an inevitable outcome.

\subsubsection{$\sigma_v = 230 \, {\rm km \, s^{-1}}$}

One more step down in $\sigma_v$ results in a solution with a more pronounced entropy inversion.  This galaxy has a total stellar mass of $10^{11.45} \, M_\odot$ and a total halo mass $M_{200} \approx 10^{12.83} \, M_\odot$.  The steady flow solution shown has $r_0 = 1$~kpc.  Stellar heating cannot push ejected stellar gas beyond $r_{\rm max} \approx 200$~kpc, and the inner cooling-flow rate reaches $\dot{M}_{\rm cf} \approx 0.1 \, M_\odot \, {\rm yr}^{-1}$.   Between 1.5 and 10~kpc, the outflow solution has a slope similar to the prediction $\alpha_K = 0.6$ from equation (\ref{eq-alpha_K_vc}), but it flattens and inverts after passing into the cooling-dominated region.  The entropy profile of the outflow solution then plunges steeply, as the cooling time becomes shorter than the flow time.  In the context of the formal solution, the entropy drop leads to a large increase in the density of the smooth outflow.  Physically, the outflow becomes convectively unstable, fragmenting into cold, dense clouds that then sink inward.  In other words, multiphase circulation is inevitable.

\subsubsection{$\sigma_v = 220 \, {\rm km \, s^{-1}}$}

Going down to $\sigma_v = 220 \, {\rm km \, s^{-1}}$ yields a steady-state solution without an inner cooling flow.  The galaxy in this panel has a total stellar mass of $10^{11.41} \, M_\odot$ and a total halo mass $M_{200} \approx 10^{12.77} \, M_\odot$.   In this potential well, stellar heating is capable of pushing ejected stellar gas out to $r_{\rm max} \approx 260$~kpc in the absence of radiative cooling, but not if the CGM pressure is significant.  Unlike the other steady solutions in Figure~\ref{schematic_K_plot_models}, this one has no stagnation point or inner cooling flow.  The difficulty is that the characteristic entropy slope of the heated outflow ($\alpha_K \approx 0.55$) is significantly smaller than the slope of the boundary between the heating- and cooling-dominated regions of the $r$--$K$ plane.  The most natural steady flows are therefore heated outflows at small radii that change into cooling outflows after crossing that boundary.  Out to $\sim 10$~kpc, the outflow solution shown has a slope similar to the prediction $\alpha_K = 0.55$ from equation (\ref{eq-alpha_K_vc}), before it flattens and inverts beyond 10~kpc.   At $r_{\rm b} = 20 \, {\rm kpc}$, the boundary pressure is $P_{\rm b} = 8.5 \times 10^{-13} \, {\rm erg \, cm^{-3}}$, about $0.02$ times the expected cosmological CGM pressure at that radius.   With this boundary condition, the steady outflow solution has $t_{\rm cool} \approx 1.5$~Gyr at 10~kpc and crosses the locus of heating-cooling equality at $r_{\rm eq} \approx 14$~kpc.  Beyond that point, radiative cooling flattens and inverts the entropy profile, preventing the outflow from going beyond 40~kpc.  Multiphase circulation within that radius is therefore unavoidable.

\section{Comparisons with X-Ray Observations}
\label{sec-ObsComparison}

This section compares X-ray observations of massive elliptical galaxies with the estimates of \S \ref{sec-BasicPicture} and some steady flow solutions similar to those in \S \ref{sec-SteadyFlow}, in order to assess the validity of the overall model and to evaluate how AGN feedback alters the steady flows that stellar mass loss would otherwise produce.  The following subsection summarizes the general trends, and the next one comments on comparisons with individual galaxies.  Readers more interested in the model's implications for quenching of star formation may wish to skip to \S \ref{sec-Implications}.

\subsection{General Trends}

\begin{figure*}[t]
\begin{center}
\includegraphics[width=6.2in, trim = 0.0in 0.1in 0.0in 0.0in]{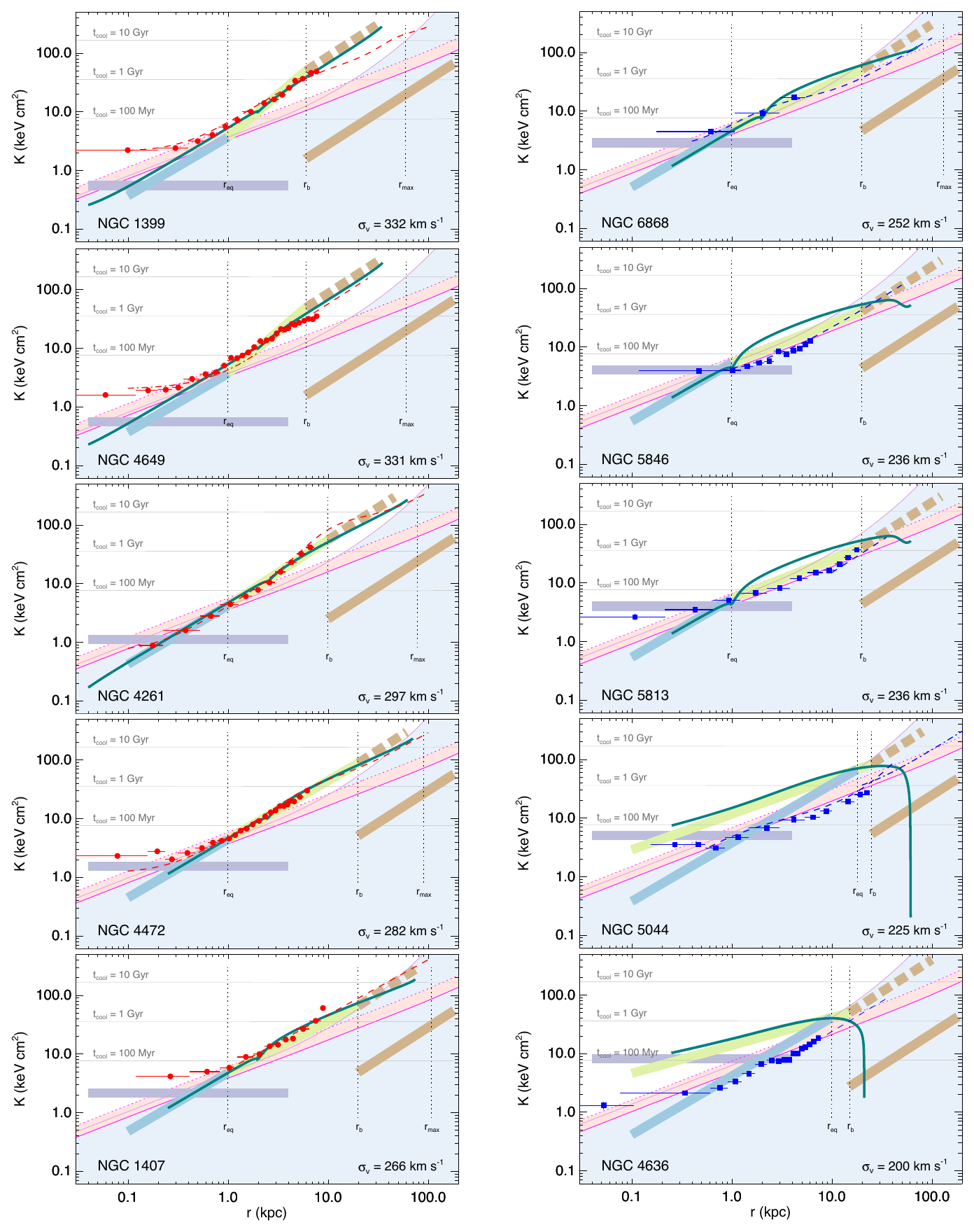} \\
\end{center}
\caption{ \footnotesize 
Comparisons of model predictions with data.  Symbols represent entropy profile data from \citet{Werner+2012MNRAS.425.2731W,Werner+2014MNRAS.439.2291W}.  Thin dashed and dotted-dashed lines represent entropy profile data from other sources described in \S \ref{sec-ObsComparison}.  All other figure elements represent the same quantities as in Figure~\ref{schematic_K_plot_models}.  The left column presents the massive ellipticals without multiphase gas beyond $\sim 1$~kpc.  The right column presents massive ellipticals that do have extended multiphase gas.  Within each column, the galaxies are arranged in order of descending stellar velocity dispersion.  All of the galaxies with $\sigma_v > 260 \, {\rm km \, ^{-1}}$ have entropy profiles that rise from $\sim 1$ to $\sim 5$~kpc more steeply than the precipitation limit, with slopes in general agreement with equation (\ref{eq-alpha_K_vc}), implying that stellar heating is driving outflows from those galaxies.  All of the galaxies with $\sigma_v < 240 \, {\rm km \, ^{-1}}$ have entropy profiles beyond $\sim 1$~kpc that track the precipitation limit at $t_{\rm cool} /t_{\rm ff} \approx 10$ and remain below $K_{\rm eq}$, implying that radiative cooling exceeds stellar heating.
\vspace*{1em}
\label{schematic_K_plot_data_steady}}
\end{figure*}

Our comparison sample of massive elliptical galaxies consists of the same 10 galaxies from \citet{Werner+2012MNRAS.425.2731W,Werner+2014MNRAS.439.2291W} analyzed by V15.  Five have extended multiphase gas beyond 1~kpc, and the other five are single phase outside of 1~kpc. 

\subsubsection{Single-phase Ellipticals}
\label{sec-SinglePhase}

Figure~\ref{schematic_K_plot_data_steady} shows that the entropy profiles of the five single-phase galaxies, which all have $\sigma_v > 260 \, {\rm km \, s^{-1}}$, are consistent with steady-state outflow solutions having $r_{\rm eq} \approx 1$~kpc and extending out to $\gtrsim 10$~kpc.  The observed entropy slopes in this radial interval agree well with the predictions of equation (\ref{eq-alpha_K_vc}), supporting the hypothesis that supernova sweeping is the primary mechanism for pushing ejected stellar gas out of these galaxies.  Section~\ref{sec-SteadyFlow} shows that the CGM pressure must be reduced to $\sim 1$\% of the cosmological value in order for supernova sweeping to succeed, and the observed profiles at $\sim 10$--100~kpc are consistent with that requirement.

Inside of 1~kpc, the entropy profiles of all but one of the single-phase ellipticals are inconsistent with a pure central cooling flow for two reasons.   First, they are flatter than the $K \propto r$ profile expected of pure cooling flows in these potential wells.  Second, they remain in the region of the $r$--$K$ plane in which stellar heating exceeds radiative cooling.  This configuration cannot persist indefinitely without AGN energy input, because the stellar source terms in regions where the pressure gradient is shallow favor a buildup of gas density over steady expansion.  The ejected stellar gas within 1~kpc is therefore trapped, meaning that its pressure and density are destined to increase and will eventually initiate a central cooling flow unless the AGN can lift the overlying gas and relieve the pressure.

Intermittent bursts of kinetic AGN feedback with a power of $\sim 10^{42} \, {\rm erg \, s^{-1}}$ can lift the ejected stellar gas and will also drive shocks capable of producing the observed central entropy plateaus (see also V15).  Thick horizontal purple lines in Figures \ref{schematic_K_plot_models} and \ref{schematic_K_plot_data_steady} show the entropy jump $\Delta K_{42}$ predicted by equation (\ref{eq-K_jets}) for a shock propagating through ambient gas at the precipitation limit and driven by $10^{42} \, {\rm erg \, s^{-1}}$ of feedback.  The observed entropy flattening is in the vicinity of those lines.  The fifth single-phase elliptical, NGC~4261, exhibits no significant central entropy flattening and is consistent with a pure cooling flow at $\sim 0.2$--1~kpc.

Accretion of hot gas onto the central black hole at the standard Bondi rate \citep{Bondi_1952MNRAS.112..195B} is able to supply $\sim 10^{42} \, {\rm erg \, s^{-1}}$ of kinetic feedback.  The Bondi accretion rate onto a black hole of mass $M_{\rm BH}$ is $\dot{M}_{\rm B} \approx 2 \pi G^2 M_{\rm BH}^2 (\mu m_p)^{5/2} (5 K_0 / 3)^{-3/2}$ in an atmosphere of constant entropy $K_0$ (V15).  Observations indicating $K_0 \approx 2 \, {\rm keV \, cm^2}$ at $\lesssim 0.5$~kpc therefore imply 
\begin{equation}
  \dot{M}_{\rm B} \approx 0.005 \, M_\odot \, {\rm yr^{-1}} 
             \left( \frac {M_{\rm BH}} {10^9 \, M_\odot} \right)^2
             K_2^{-3/2}
             \; \; ,
\end{equation}
where $K_2 \equiv K_0 / (2 \, {\rm keV \, cm^2})$.  At this rate, conversion of $\sim 1$\% of the accreting rest-mass energy into kinetic feedback energy is sufficient to produce $\sim 10^{42} \, {\rm erg \, s^{-1}}$.  However, considerably more time-averaged power is required to lift the CGM.

Occasional episodes of chaotic cold accretion can provide the power needed to lift the CGM when the ambient central pressure becomes large enough to lower the central cooling time to $t_{\rm cool} \approx 10 t_{\rm ff}$.  Those episodes tend to boost accretion by a factor as great as $\sim 10^2$ over the ambient Bondi accretion rate \citep{Gaspari+2013MNRAS.432.3401G}, raising the kinetic feedback power to $\sim 10^{44} \, {\rm erg \, s^{-1}}$, assuming a $\sim 1$\% conversion efficiency.  They are ultimately limited by the maximum cooling-flow rate within $\sim 1$~kpc, which is $\sim 0.5 \, M_\odot \, {\rm yr}^{-1}$ in all of these galaxies.  NGC~4261 appears to be a galaxy that is currently experiencing chaotic cold accretion because it has $t_{\rm cool} / t_{\rm ff} \lesssim 10$ at $\lesssim 0.5$~kpc, a multiphase medium at $\lesssim 100$~pc that includes a cold gaseous disk, and is producing $\sim 10^{44} \, {\rm erg \, s^{-1}}$ of kinetic feedback power.

\newpage 

\subsubsection{Multiphase Ellipticals}

The four multiphase galaxies with $\sigma_v < 240 \, {\rm km \, s^{-1}}$ are inconsistent with homogeneous steady flow solutions.  Instead, they track the precipitation limit at $t_{\rm cool}/t_{\rm ff} \approx 10$ (see also Voit et al. 2015).  This result is consistent with the finding of \S \ref{sec-SteadyFlow} that galaxies with $\sigma_v \lesssim 240 \, {\rm km \, s^{-1}}$ are prone to multiphase circulation and precipitation.  The confining CGM pressure at 10--20~kpc is 3\%--7\% of the cosmological expectation, several times greater than the confining pressure around the single-phase galaxies.  Three of the four exhibit entropy profile flattening at $\lesssim 1$~kpc near the level expected from intermittent AGN feedback with a kinetic power of $\sim 10^{42} \, {\rm erg \, s^{-1}}$.  The fourth (NGC~4636) is consistent with a pure cooling flow from $\sim 0.5$ to $\sim 3$~kpc, but its entropy profile is flatter than $K \propto r$ inside of 0.5~kpc.  The remaining multiphase elliptical (NGC~6868, $\sigma_v = 252 \, {\rm km \, s^{-1}}$), is sparsely sampled but appears to be intermediate between the supernova-sweeping and precipitation-limited cases.

Comparing stellar heating with radiative cooling in the four galaxies with $\sigma_v < 240 \, {\rm km \, s^{-1}}$ shows that cooling dominates outside of $\sim 1$~kpc with increasing significance toward larger radii.  An outflow driven by stellar heating is therefore not plausible.  In these galaxies, the primary driver of ejected stellar gas out of the galaxy must be the AGN.   Kinetic feedback with a time-averaged power of $\sim 10^{41.5} \, {\rm erg \, s^{-1}}$ is necessary (see \S \ref{sec-TheValve}) and is consistent with the level of central entropy flattening observed in NGC~5846, NGC~5813, and NGC~5044.  However, the center of NGC 4636 may be trending toward a pure cooling-flow state destined to trigger a more powerful AGN outburst, similar to the one in NGC~4261.  Galaxies like these may be intermittently switching between Bondi accretion and chaotic cold accretion.

Outflows driven by kinetic feedback in the galaxies with $\sigma_v < 240 \, {\rm km \, s^{-1}}$ are unlikely to be uniform and homogeneous, given that $t_{\rm cool} / t_{\rm ff} \approx 10$ within much of the galaxy.  Nearly adiabatic uplift of such a medium lowers $t_{\rm cool} / t_{\rm ff}$ in the uplifted gas, making making it highly susceptible to multiphase condensation \citep{Revaz_2008A&A...477L..33R,LiBryan2014ApJ...789..153L,McNamara_2016ApJ...830...79M,Voit_2017_BigPaper}.  Simultaneously, anisotropic AGN energy input tends to drive convection and turbulence, further destabilizing the medium \citep{TaborBinney1993MNRAS.263..323T,ps05,Gaspari+2013MNRAS.432.3401G,Meece_2017ApJ...841..133M,Voit_2017_BigPaper}.  Both factors are likely to be responsible for the extended multiphase gas observed in these galaxies.  Yet their observed star formation rates are $\lesssim 0.1 \, M_\odot \, {\rm yr}^{-1}$ \citep{Werner+2014MNRAS.439.2291W}, yielding specific star formation rates $\lesssim 10^{-12} \, {\rm yr}^{-1}$, making them all formally ``quenched."

\subsection{Comments on Individual Galaxies}

\subsubsection{NGC 1399}

The upper left panel of Figure~\ref{schematic_K_plot_data_steady} plots observations of NGC 1399, the central galaxy of the Fornax Cluster, on an entropy profile graph similar to the ones in Figure~\ref{schematic_K_plot_models}.   Red symbols show an entropy profile derived from {\em Chandra} entropy observations by \citet{Werner+2014MNRAS.439.2291W}, and a red dashed line shows an entropy profile derived from the density and temperature fits of \citet{Paolillo_2002ApJ...565..883P} to ROSAT data.  The steady outflow model shown is specified by $\sigma_v = 332 \, {\rm km \, s^{-1}}$ and $r_0 = 2$~kpc and is a good description of the data from $\sim 0.5$ to at least 8~kpc.  At larger radii, the observed $K(r)$ profile becomes shallower than the steady outflow solution, suggesting that CGM confinement of the outflow begins near $\sim 10$~kpc.  We have placed the line marking $r_{\rm b}$ at 6~kpc, where there is an inflection in the observed entropy profile.  Inside of 0.5~kpc, the entropy profile flattens at $\sim 2 \, {\rm keV \, cm^2}$, consistent with AGN shock heating by kinetic outflows ranging up to a few times $10^{42} \, {\rm erg \, s^{-1}}$.

\subsubsection{NGC 4649}

In the panel for NGC 4649, the central galaxy of a Virgo Cluster subgroup, red symbols show an entropy profile derived by \citet{Werner+2014MNRAS.439.2291W} from {\em Chandra} observations.  A red dashed line from 1 to 25~kpc shows an entropy profile derived from the densities and temperatures observed by \citet{Randall_N4649_2004ApJ...600..729R}, and a dotted-dashed red line from 0.1 to 2~kpc shows the entropy profile fit of \citet{Humphrey_2006ApJ...646..899H}.  The steady outflow model shown is specified by $\sigma_v = 331 \, {\rm km \, s^{-1}}$ and $r_0 = 2$~kpc.  It is a good description of the data from $\sim 0.5$ to $> 20$~kpc, but the Werner et al. observations suggest an inflection in the entropy profile near 6~kpc, where we have placed the $r_{\rm b}$ marker.   Inside of 0.5~kpc, the observed $K(r)$ profile becomes shallower than the outflow solution, flattening slightly below $2 \, {\rm keV \, cm^2}$,  consistent with intermittent shock heating by AGN kinetic outflows with $\sim 10^{42} \, {\rm erg \, s^{-1}}$.

\subsubsection{NGC 4261}

The case of NGC~4261, a central group galaxy, is perhaps the most revealing. Red symbols represent {\em Chandra} observations by \citet{Werner+2014MNRAS.439.2291W}, and a dashed red line shows a fit to observations by \citet{Humphrey_2009ApJ...703.1257H}.  From 0.1 to 10~kpc, the data points are consistent with a steady flow solution determined by $\sigma_v = 297 \, {\rm km \, s^{-1}}$ and $r_0 = 2.5$~kpc.  Within this stagnation radius, the inner cooling-flow rate reaches $\approx  0.5 \, M_\odot \, {\rm yr^{-1}}$.  As the flow moves inward, the ambient $t_{\rm cool}/t_{\rm ff}$ ratio declines, ultimately dropping below $t_{\rm cool}/t_{\rm ff} \approx 10$ at $r \lesssim 0.5 \, {\rm kpc}$, indicating that the inflow becomes increasingly prone to multiphase condensation (V15).  This multiphase inflow presumably supplies gas to the central dusty disk, which has a radius of $\sim 100$~pc \citep{Jaffe_N4261_1996ApJ...460..214J}.  Accretion of the inflowing gas onto the central black hole is therefore capable of providing the current AGN power, assuming a conversion efficiency $\gtrsim 1$\% from rest-mass energy to kinetic feedback power.
  
Comparing stellar heating with radiative cooling in NGC 4261 shows that heating dominates outside of 1 kpc and should therefore drive an outflow at larger radii.   The data are consistent with an outflow driven by stellar heating having the entropy profile slope predicted by equation (\ref{eq-alpha_K_vc}) at radii of $\sim 2$--10 kpc.  The energy required to lift the CGM in the potential well of NGC 4261 is $\sim f_{\rm b} M_{200}  \sigma_v^2 \sim 10^{60.7} \, {\rm erg}$.   This amount of energy can plausibly be supplied by the AGN if its kinetic power output has been close to the current $\sim 10^{44} \, {\rm erg \, s^{-1}}$ for several Gyr of its history.

\subsubsection{NGC 4472}

Observations of NGC~4472, which dominates its own subgroup of the Virgo Cluster, are consistent with a steady flow solution from $\sim 0.5$ to beyond 20~kpc.  Red symbols show data from Werner et al., the dashed red line shows a fit from \citet{Humphrey_2009ApJ...703.1257H}, and the teal line represents a steady flow model specified by $\sigma_v = 282 \, {\rm km \, s^{-1}}$ and $r_0 = 2.5$~kpc.  Within $\sim 0.5$~kpc, the observed entropy profile departs from the steady flow model, flattening near $2 \, {\rm keV \,cm^2}$, in the vicinity of the horizontal purple line indicating the entropy jumps of intermittent shocks driven by $\sim 10^{42} \, {\rm erg \, s^{-1}}$ of AGN power.  Stellar heating exceeds radiative cooling at all radii.  Here V15 found $\min(t_{\rm cool}/t_{\rm ff}) \approx 20$, indicating that the atmosphere in its current configuration does not condense into cold clouds that fuel the central black hole.  Given the observed core entropy level and an observed black hole mass of $\approx 2.5 \times 10^9 \, M_\odot$  \citep{KormendyHo2013ARAA..51..511K}, the resulting Bondi accretion rate of hot gas onto the black hole is $\sim 0.05 \, M_\odot \, {\rm yr}$, which yields $\sim 10^{43} \, {\rm erg \, s^{-1}}$ of kinetic power and suggests a conversion efficiency of $\sim 1$\%.

\subsubsection{NGC 1407}

Red symbols in the panel for NGC~1407, a central group galaxy, show the Werner et al. observations, and the dashed red line is based on the fits made by \citet{Su_N1407_2014ApJ...786..152S} to their observations.  From $\sim 1$ to 8 kpc, the data generally agree with the steady flow solution having $\sigma_v = 266 \, {\rm km \, s^{-1}}$ and $r_0 = 2$~kpc, as well as the simple power law predicted by equation (\ref{eq-alpha_K_vc}) and shown by the thick green line.   From 10 to 20 kpc, where the temperature profile peaks, there is a bump in the entropy profile.  At larger radii, the entropy profile is more consistent with a cosmological slope $K \propto r^{1.1}$ than with the steady outflow solution, and so we have placed the $r_{\rm b}$ marker at 10~kpc.  Inside of $\sim 1$~kpc, flattening of the entropy profile suggests intermittent shock heating, although the inner region is not as well resolved as in some of the other galaxies.

\subsubsection{NGC 6868}

Blue symbols in the panel for NGC~6868, one of two large galaxies in the Telescopium group, represent the three data points from \citet{Werner+2014MNRAS.439.2291W}.  A teal line shows a steady flow solution with $\sigma_v = 252 \, {\rm km \, s^{-1}}$ and $r_0 = 2$~kpc.  The data are sparse but consistent with the predicted entropy slope ($\alpha_K = 0.7$) from equation (\ref{eq-alpha_K_vc}).  This slope makes the entropy profile of NGC~6868 nearly parallel to the precipitation limit, and it tracks near $t_{\rm cool} / t_{\rm ff} \approx 20$ (V15).  \citet{Juranova_2019MNRAS.484.2886J} have recently suggested that rotation makes the gas in this galaxy especially prone to precipitation.

\subsubsection{NGC 5846}

The central group galaxy NGC~5846, has $\sigma_v < 240 \, {\rm km \, s^{-1}}$ and the blue symbols representing \citet{Werner+2014MNRAS.439.2291W} entropy observations are inconsistent with steady homogenous flow.  Instead, they track the precipitation limit at $t_{\rm cool} / t_{\rm ff} \approx 10$ from $\sim 1$ to $\gtrsim 6$ kpc.  Beyond that point, the the dashed blue line shows the entropy profile observed by \citet{Paggi_2017ApJ...844....5P}, which gradually steepens until it matches the cosmological slope near 20~kpc, where we have placed the $r_{\rm b}$ marker.  At that radius, the CGM pressure is $\sim 3$\% of the cosmological pressure expected without radiative cooling or feedback.  Inside of 1~kpc, the entropy profile appears to flatten at a level consistent with $\sim 10^{42} \, {\rm erg \, s^{-1}}$ of intermittent kinetic feedback power but is not well resolved.

\subsubsection{NGC 5813}

The largest galaxy in its subgroup of the NGC~5846 group, NGC~5813 has a velocity dispersion nearly identical to NGC~5846 and a nearly identical entropy profile, with blue symbols representing \citet{Werner+2014MNRAS.439.2291W} data and a blue dashed line representing \citet{Randall_N5813_2015ApJ...805..112R} data.  It tracks the precipitation limit from 1 to 10 kpc and is inconsistent with steady homogeneous flow.  Beyond 10~kpc, its entropy profile steepens to a slope similar to the $K \propto r^{1.1}$ slope produced by cosmological structure formation, but its entropy normalization at 20~kpc is greater by a factor of $\sim 9$, corresponding to a CGM pressure normalization of $\sim 3$\% of the cosmological expectation.  Inside of 1~kpc, the entropy profile becomes flatter than $K \propto r^{2/3}$, leveling at $\sim 3 \, {\rm keV \, cm^2}$, again consistent with $\sim 10^{42} \, {\rm erg \, s^{-1}}$ of intermittent kinetic feedback power.

\subsubsection{NGC 5044}

The entropy profile of NCG 5044, a central group galaxy, is similar in many ways to those of NGC 5846 and NGC 5813 but tracks the precipitation limit to even greater radii, as shown by the blue symbols representing \citet{Werner+2014MNRAS.439.2291W} observations.  A dotted-dashed line beyond 20~kpc shows a fit to observations by \citet{David_N5044_2017ApJ...842...84D}, which attains a cosmological ($K \propto r^{1.1}$) slope outside of the $r_{\rm b}$ marker at 25~kpc.  Here the pressure is $\sim 7$\% of the cosmological expectation.  A dashed blue line at smaller radii shows a fit to observations by \citet{David_2009ApJ...705..624D}.  Inside of 1~kpc, the profile levels off near $3 \, {\rm keV \, cm^2}$, indicating $\sim 10^{42} \, {\rm erg \, s^{-1}}$ of intermittent kinetic feedback power.
 
\subsubsection{NGC 4636}

The central group galaxy NGC~4636 has an entropy profile that tracks the precipitation limit from 0.5 to 10 kpc. The blue dashed line represents data from \citet{Trinchieri_1994ApJ...428..555T} and is consistent with a cosmological slope outside of 10~kpc.  The \citet{Werner+2014MNRAS.439.2291W} observations depicted by blue symbols are in the vicinity of the precipitation limit from 0.5 to 8 kpc but are also consistent with a pure cooling flow between 0.5 and 2 kpc.  At smaller radii, the entropy profile flattens relative to the the $K \propto r^{2/3}$ precipitation-limited profile but reaches $\sim 1 \, {\rm keV \, cm^2}$ inside of 100~pc, considerably below the level expected from $\sim 10^{42} \, {\rm erg \, s^{-1}}$ of intermittent kinetic feedback power.  There are several possible explanations for this low central entropy level: (1) the time-averaged kinetic AGN power has been $\sim 10^{41} \, {\rm erg \, s^{-1}}$ for the last $\sim 100$~Myr, (2) the AGN power has been highly collimated, as in NGC~4261, and has penetrated to $\gg 1$~kpc without dissipating much power; or (3) the AGN power has been too weak to balance cooling for the last $\sim 100$~Myr.  In this last case, a cooling catastrophe is imminent, as suggested by the entropy profile between 0.5 and 2 kpc, and will soon trigger a strong feedback episode.

\section{Implications for Quenching}
\label{sec-Implications}

The black hole feedback valve model presented here potentially solves some important puzzles in galaxy evolution.  For example, AGN feedback is energetically necessary to quench star formation in massive galaxies, but quenching itself appears to be most closely related to a galaxy's central stellar velocity dispersion \citep{Wake_2012ApJ...751L..44W,Teimoorinia_2016MNRAS.457.2086T,Bluck_2016MNRAS.462.2559B,Bluck_2020MNRAS.492...96B} or, equivalently, to the surface density of stars within the central 1~kpc \citep{Bell_2008ApJ...682..355B,Franx_2008ApJ...688..770F,Fang_2013ApJ...776...63F,vanDokkum_2015ApJ...813...23V,Woo_2015MNRAS.448..237W,Whitaker_2017ApJ...838...19W}.   The AGN feedback must somehow be related to galactic structure, but which way does the arrow of causality point?  Does galactic structure determine the rate of black hole growth, or does AGN feedback shape galactic structure?   Also, why is AGN feedback more effective at quenching star-formation in massive galaxies than in smaller ones, even though far more energy is required to offset radiative cooling, alleviate CGM pressure confinement, and prevent accumulation of ejected stellar gas?

According to the model, AGN feedback is responding to galactic structure.  A galaxy's ability to remain quenched for long time periods depends most critically on the entropy gradient of its ambient gas.  A steep entropy slope ($\alpha_K > 2/3$) strongly inhibits widespread star formation for two reasons.
\begin{enumerate}

\item The corresponding gas density slope is steeper than $n_e \propto r^{-1}$, minimizing the local ratio of stellar heating to radiative cooling at small radii ($\lesssim 1$~kpc) and allowing local stellar heating to exceed local radiative cooling at larger radii ($\approx 1$--10~kpc).  

\item The $t_{\rm cool} / t_{\rm ff}$ ratio rises with radius.  Multiphase condensation therefore happens primarily near the central black hole, where $t_{\rm cool} / t_{\rm ff}$ is minimized, and is suppressed by buoyancy effects at larger radii. 

\end{enumerate}
A fundamentally important consequence is that the central stellar mass density of a galaxy should be self-limiting, because equation (\ref{eq-alpha_K_vc}) predicts that growth in $\alpha_K$ should accompany growth in $\sigma_v$.  When $\sigma_v$ becomes large enough, only the region within $\sim 1$~kpc of the central black hole can persist in a state with $\min(t_{\rm cool} / t_{\rm ff}) \sim 10$, resulting in episodes of chaotic cold accretion that intermittently supercharge AGN feedback while strongly limiting star formation elsewhere.  

Figure~\ref{schematic_K_plot_data_steady} indicates that present-day elliptical galaxies with $\sigma_v > 240 \, {\rm km \, s^{-1}}$ have attained such a state, which requires AGN feedback to have lowered the confining CGM pressure by a factor of $\sim 10^2$ relative to expectations from cosmological structure formation.  However, the other galaxies in Figure~\ref{schematic_K_plot_data_steady} are also quenched and have $\sigma_v = 200$ to $237 \, {\rm km \, s^{-1}}$, suggesting that the dependence of quenching on $\sigma_v$ is a continuous transition rather than a step function at the critical value.  Also, the model predicts that the critical value of $\sigma_v$ should be time-dependent, because the ratio of SN Ia heating to stellar mass loss does not remain constant with time.  The remainder of this section discusses these issues, along with some other implications of the black hole feedback valve model.

\subsection{Maximum Stellar Velocity Dispersion}
\label{sec-UpperLimit}

Equation (\ref{eq-alpha_K_vc}) predicts that cooling and condensation of ambient gas should become increasingly concentrated toward the center of a galaxy as its stellar velocity dispersion $\sigma_v$ increases.  The result was derived for quiescent stellar populations but has more general applications.  For example, consider an actively star forming galaxy in which explosions of massive stars (SNe II) are driving an outflow at a rate $\eta$ times the star formation rate $\dot{M}_*$.  If star formation results in $10^{51} \, {\rm erg}$ of supernova energy per $100 \, M_\odot$ of star formation, then the specific thermal energy of a supernova-heated outflow is no greater than $\epsilon_* \approx 3 \, \eta^{-1} \, {\rm keV} / \mu m_p$, and possibly much less if radiative losses are substantial.  The critical velocity dispersion at which $\alpha_K \approx 2/3$ in the outflow is then no greater than
\begin{equation}
  \sigma_v \approx 300 \, \eta^{-1/2} \, {\rm km \, s^{-1}}
  \label{eq-sigma_v_nugget}
  \; \; . 
\end{equation}
Steady supernova-driven flows in galaxies that exceed this limit are cooling-dominated at small radii and become increasingly focused on the central black hole as $\sigma_v$ rises.  Once $\alpha_K$ exceeds 2/3, the black hole feedback valve described in \S \ref{sec-TheValve} responds by lowering the confining CGM pressure until $t_{\rm cool} / t_{\rm ff} > 10$ outside of the central kiloparsec.  

A galaxy in this state remains quenched indefinitely because buoyancy prevents multiphase condensation of the ambient medium, except near the central black hole or during eruptions of AGN feedback that lift large quantities of low-entropy gas to greater altitudes.  This limiting effect of the black hole feedback valve on $\sigma_v$ should also be present in numerical simulations of galaxy evolution that implement AGN feedback in the form of bipolar jets capable of thermalizing their energy in the CGM after drilling through the ambient medium out to $\gtrsim 10$~kpc.  High spatial resolution is necessary because the jets need to be much narrower than a kiloparsec at the base in order to pass through the central few kiloparsecs (as in NGC~4261) without completely disrupting the ambient gas there.  

A dramatic demonstration of what happens without such a feedback mechanism can be found in \citet{Keller_2016MNRAS.463.1431K}.  Figure~4 of that paper presents rotation curves for a set of simulated massive galaxies with efficient superbubble feedback but no AGN feedback.  Those galaxies fall into two distinct subsets.  One has flat rotation curves with $\max (v_{\rm c}) < 250 \, {\rm km \, s^{-1}}$, while the other has rotation curves with a sharp peak at $\lesssim 1$~kpc at which $450 \, {\rm km \, s^{-1} } \lesssim v_{\rm c} \lesssim 700 \, {\rm km \, s^{-1} }$.  Apparently, the galaxies with centrally peaked rotation curves experienced episodes of centrally focused cooling and runaway star formation after $\max (v_{\rm c})$ exceeded a critical value between 250 and $450 \, {\rm km \, s^{-1}}$ that corresponds to $180 \, {\rm km \, s^{-1}} \lesssim \sigma_v \lesssim 320 \, {\rm km \, s^{-1}}$.  If kinetic AGN feedback had been enabled in these simulations, centrally focused cooling would instead have shut down star formation shortly following the onset of the runaway, thereby preventing the central stellar velocity dispersion from greatly exceeding $300 \, {\rm km \, s^{-1}}$.

This limiting effect of AGN feedback effect on $\sigma_v$ does indeed show up in cosmological simulations of massive galaxies by \cite{Choi_2018ApJ...866...91C}.  Without AGN feedback, the central concentration of star formation in their galaxies causes the central stellar mass density to grow to a level corresponding to $\sigma_v > 500 \, {\rm km \, s^{-1}}$ by $z \approx 0$.  In those same galaxies, implementation of a kinetic AGN feedback mechanism limits the central stellar mass density to an equivalent stellar velocity dispersion in the range $250 \, {\rm km \, s^{-1}} \lesssim \sigma_v \lesssim 400 \, {\rm km \, s^{-1}}$.

Observations of ``red nugget" galaxies \citep[e.g.,][]{Damjanov_2009ApJ...695..101D} provide additional support for this limiting mechanism.  That population of galaxies became quenched early in the history of the universe (at $z \gtrsim 2$), with a particularly small size and large stellar mass density.  Forming them required a highly dissipative process to concentrate much of the star forming gas within a volume $\sim 1$~kpc in radius.  Star formation then ceased, presumably because of an episode of strong AGN feedback, when the red nugget reached a stellar velocity dispersion in the range $250 \, {\rm km \, s^{-1}} \lesssim \sigma_v \lesssim 400 \, {\rm km \, s^{-1}}$ \citep[e.g.,][]{delaRosa_2016MNRAS.457.1916D}, consistent with equation (\ref{eq-sigma_v_nugget}).  High-resolution X-ray observations of red nuggets at $z > 1$ are currently not feasible, but examples of analogous galaxies in the low-redshift universe have $\alpha_K > 2/3$ and $t_{\rm cool} / t_{\rm ff} > 10$ \citep{Werner_2018MNRAS.477.3886W,BuoteBarth_2019ApJ...877...91B}, also as expected from the model.

\subsection{Dependence of Quenching on $\sigma_v$}
\label{sec-Dependence}

A velocity dispersion $\sigma_v > 240 \, {\rm km \, s^{-1}}$ appears to be a sufficient condition for the quenching of present-day ellipticals through the black-hole feedback valve mechanism, but it is not a necessary condition.  For example, Figure~\ref{schematic_K_plot_data_steady} shows several elliptical galaxies with $\sigma_v < 240 \, {\rm km \, s^{-1}}$ and specific star-formation rates of $\sim 10^{-12} \, {\rm yr}^{-1}$.  The presence of extended multiphase gas in those galaxies suggests that AGN feedback cannot completely suppress multiphase condensation but is still sufficiently well coupled with the CGM to strongly suppress star formation.  However, the fraction of galaxies that are quenched is observed to decline with decreasing $\sigma_v$, implying that AGN feedback is less well coupled to the CGM in smaller galaxies. 
 
Analyses of large galaxy samples from the Sloan Digital Sky Survey show that quenching correlates more closely with $\sigma_v$ than with any other galactic property \citep{Wake_2012ApJ...751L..44W,Teimoorinia_2016MNRAS.457.2086T,Bluck_2016MNRAS.462.2559B,Bluck_2020MNRAS.492...96B}.  When a quenched state is defined to be a specific star formation rate less than 10\% of the average among star-forming galaxies of similar mass \cite[e.g.,][]{Bluck_2014MNRAS.441..599B}, the fraction of both central and satellite galaxies that qualify as quenched is $\gtrsim 90$\% for $\sigma_v > 240 \, {\rm km \, s^{-1}}$, with no apparent dependence on $\sigma_v$.  Among central galaxies with $\sigma_v < 240 \, {\rm km \, s^{-1}}$, the quenched fraction continuously declines to $\sim 25$\% at $\sigma_v = 100 \, {\rm km \, s^{-1}}$  \citep{Bluck_2016MNRAS.462.2559B}.  Among satellite galaxies the quenched fraction also declines below $\sigma_v = 240 \, {\rm km \, s^{-1}}$, but not as steeply, down to $\sim 50$\% at $\sigma_v = 100 \, {\rm km \, s^{-1}}$.

The black-hole feedback valve model outlined in this paper is not sophisticated enough to make quantitative predictions for this dependence of quenched fraction on $\sigma_v$ or to model environmental effects.  It will therefore need to be tested with numerical simulations employing feedback algorithms that produce galaxies similar to those shown in Figure~\ref{schematic_K_plot_data_steady}.   \citet{Wang_2019MNRAS.482.3576W} already made progress by simulating galaxies resembling NGC~4472 and NGC~5044.  In their simulations, AGN feedback keeps the galaxy resembling NGC~4472 quenched for several Gyr without producing extended multiphase gas, while the galaxy resembling NGC~5044 develops a persistent multiphase medium but still remains quenched.  It will be intriguing to see how the same algorithms play out in simulated galaxies with $\sigma_v < 200 \, {\rm km \, s^{-1}}$.

\begin{figure*}[t]
\begin{center}
\includegraphics[width=6.5in, trim = 0.0in 0.0in 0.0in 0.0in]{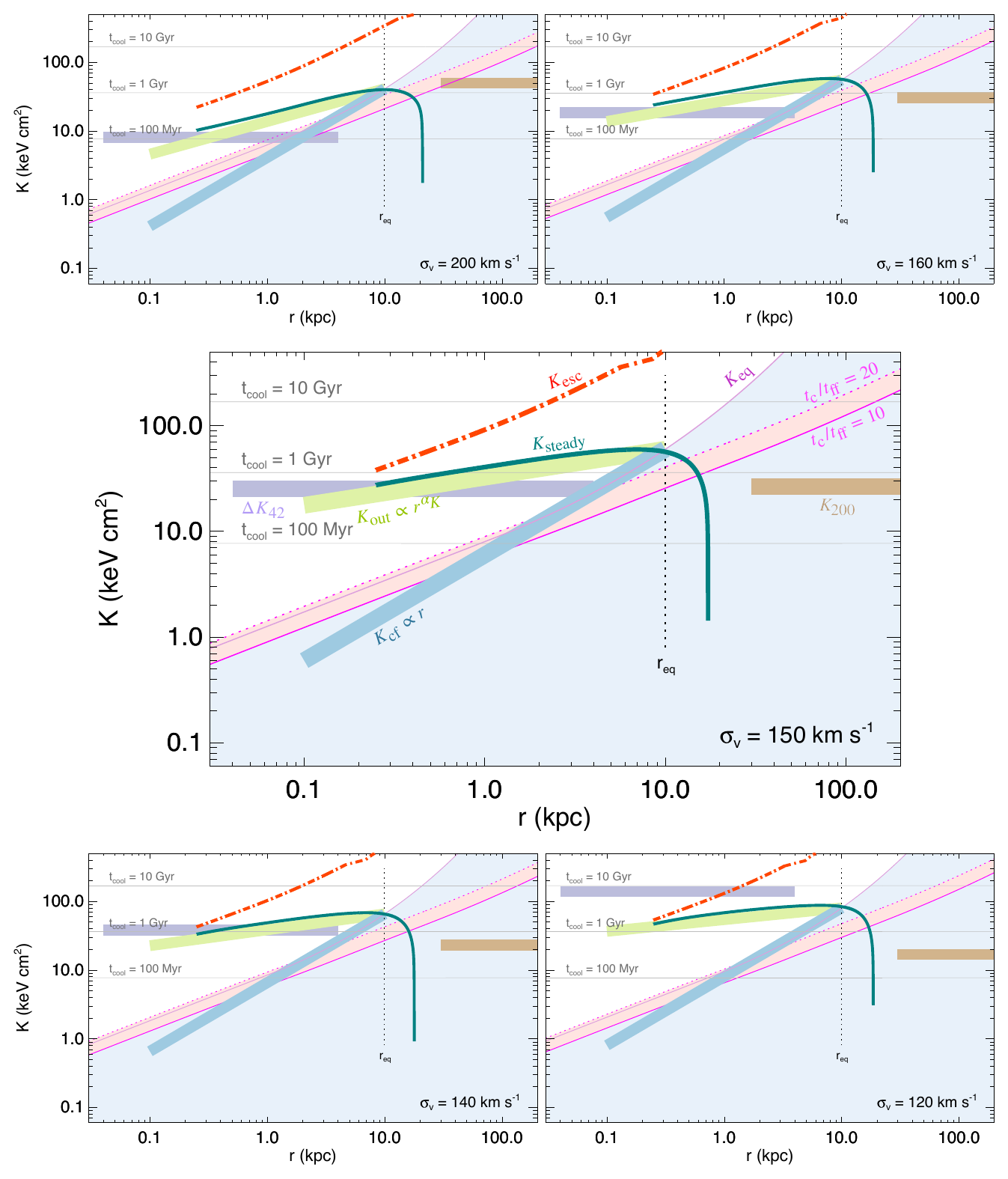} \\
\end{center}
\caption{ \footnotesize 
Characteristic entropy profiles for generic galaxies with $120 \, {\rm km \, s^{-1}} \lesssim \sigma_v \lesssim 200 \,  {\rm km \, s^{-1}}$.  All figure elements have the same meanings as in Figure~\ref{schematic_K_plot_models}, except for the horizontal tan lines from 30 to 200 kpc, which show the cosmological CGM entropy $K_{200}$ typically generated by accretion shocks (see equation \ref{eq-K200}).  A galaxy in a precipitation-limited state will have an entropy profile in the vicinity of the magenta lines.  As $\sigma_v$ declines, the horizontal purple line corresponding to $10^{42} \, {\rm erg \, s^{-1}}$ of feedback rises, while the tan line indicating cosmological CGM entropy drops.  Feedback in lower-mass galaxies therefore tends to produce entropy inversions that promote multiphase circulation and stimulate condensation of clouds capable of fueling star formation.  Consequently, AGN feedback is less effective at quenching of star formation in galaxies with lower $\sigma_v$.  However, stripping of the CGM around those galaxies can allow SN Ia to drive transonic winds corresponding to the dotted-dashed orange lines (labeled $K_{\rm esc}$ in the central panel), which have entropy profiles that enable buoyancy to suppress condensation and star formation.
\vspace*{1em}
\label{circflow_K_plot_models}}
\end{figure*}

Generically, we expect this AGN feedback mechanism to be less effective at quenching smaller galaxies because the CGM entropy gradient in those galaxies is less able to inhibit multiphase condensation during feedback bursts, for the reasons shown in Figure~\ref{circflow_K_plot_models}.  Solid magenta lines in that figure indicate the precipitation limit at $t_{\rm cool} / t_{\rm ff} \approx 10$, along which ambient gas is marginally susceptible to multiphase condensation.  The X-ray observations show that the central entropy profiles of early-type galaxies rarely, if ever, fall below that limit.  Data of sufficient quality to derive resolved entropy profiles like those in Figure~\ref{schematic_K_plot_data_steady} are available for only a few galaxies in this range of $\sigma_v$ \citep[e.g.,][]{Babyk_2018ApJ...862...39B}.  However, larger samples show that X-ray luminosity from within the effective radius of an early-type galaxy does not exceed the limit imposed by the condition $\min (t_{\rm cool} / t_{\rm ff}) \gtrsim 10$ and is more often consistent with $\min (t_{\rm cool} / t_{\rm ff}) \approx 20$--30 \citep{Goulding_2016ApJ...826..167G,Voit2018_LX-T-R}.

Given that condition, Figure~\ref{circflow_K_plot_models} shows that bursts of AGN feedback tend to raise the central entropy level above the cosmological CGM entropy level in potential wells with $\sigma_v \lesssim 150 \, {\rm km \, s^{-1}}$.  Horizontal purple lines in the figure indicate the entropy jump produced by $10^{42} \, {\rm erg \, s^{-1}}$ of kinetic feedback, as given by equation (\ref{eq-K_jets}).  Horizontal tan lines indicate the CGM entropy scale produced by cosmological structure formation.  Those latter lines mark
\begin{equation}
  K_{200} \equiv kT_\phi \left( \frac {200 f_{\rm b} \rho_{\rm cr}} {\mu_e m_p} \right)^{-2/3}
  \label{eq-K200}
\end{equation}
and represent the typical CGM entropy level resulting from accretion shocks \citep{Voit+03,Voit_2005RvMP...77..207V}.  If the tan line is below the purple line, then feedback near the center of the potential well produces bubbles of high-entropy gas that buoyantly rise through lower-entropy CGM gas.   This configuration is unstable and promotes multiphase condensation of the lower-entropy gas \citep[e.g.,][]{McNamara_2016ApJ...830...79M,Voit_2017_BigPaper}.  Furthermore, even if the tan line is slightly above the purple line, the corresponding entropy gradient remains shallow, meaning that small CGM disturbances are able to promote multiphase condensation.  In Figure~\ref{circflow_K_plot_models}, the tan line falls below the purple line at $\sigma_v \lesssim 140 \, {\rm km \, s^{-1}}$, for which the generic galaxy model gives $M_* \lesssim 10^{10.8} \, M_\odot$ and $M_{200} \lesssim 10^{12.2} \, M_\odot$.  This is indeed the boundary below which most of the universe's star formation now occurs, but the role of entropy gradients in establishing it needs to be explored more quantitatively with high-resolution simulations of cosmological galaxy evolution.

\citet{Bower_2017MNRAS.465...32B} have proposed a similar explanation for how AGN feedback can cause star-formation quenching in halos with $\gtrsim 10^{12} \, M_\odot$.  However, they emphasized the role of halo mass instead of $\sigma_v$.  Both proposed explanations recognize that hot bubbles produced at small radii by feedback will buoyantly rise to large radii in halos with $\lesssim 10^{12} \, M_\odot$. That happens because the entropy produced by accretion shocks in low-mass galaxies tends to be smaller than the entropy produced closer to the galaxy by feedback heating.  \citet{Bower_2017MNRAS.465...32B} argued that quenching happens in halos with $\gtrsim 10^{12} \, M_\odot$ because the higher cosmological CGM entropy prevents supernova-heated bubbles from rising.  The resulting buildup of galactic gas then triggers AGN feedback, causing an eruption of energy that heats the CGM and quenches star formation. 

Motivated by the fact that quenching is observed to correlate much more directly with $\sigma_v$ than with halo mass, we argue here that the role of buoyancy is more subtle.  In the black hole feedback valve model, both supernova and AGN feedback can inflate high-entropy bubbles that add heat to the CGM and regulate star formation in halos with $\lesssim 10^{12} \, M_\odot$.  However, those bubbles fail to quench star formation in lower-mass galaxies because they produce large-scale entropy inversions that promote multiphase condensation.  

In order to be effective at long-term quenching of star formation, AGN feedback must maintain both $t_{\rm cool} / t_{\rm ff} > 10$ and a positive entropy gradient sufficient for buoyancy to suppress multiphase condensation.  We have shown that AGN feedback is able to do so when the depth of the galactic potential well is comparable to the specific energy of supernova heating.  As the entropy gradient of a supernova-heated outflow rises with increasing $\sigma_v$ (see the green and teal lines in Figure~\ref{circflow_K_plot_models}), the ability of buoyancy to suppress condensation increases.  And the entropy slope becomes great enough for quenching to be inevitable when $\sigma_v \gtrsim 240 \, {\rm km \, s^{-1}}$.  Halo mass plays a secondary role because it determines the CGM entropy level produced by accretion shocks, which enhances the effects of buoyancy if the cosmological entropy is great enough.

\subsection{Redshift Evolution of the Critical $\sigma_v$}
\label{sec-Evolution}

One observationally testable prediction of the black hole feedback valve model is that the $\sigma_v$ scale for quenching should be greater earlier in time because the specific energy of stellar ejecta ($\epsilon_*$) from a younger stellar population is greater.  Section~\ref{sec-UpperLimit} showed that the critical velocity dispersion is likely to be greater than $240 \, {\rm km \, s^{-1}}$ in an actively star-forming population, but even in a quiescent stellar population, $\epsilon_*$ should be greater earlier in time.  That is because the specific SN Ia rate is $\propto t^{-1.3}$ in massive ellipticals \citep[e.g.,][]{FriedmannMaoz_2018MNRAS.479.3563F}, while the specific stellar mass-loss rate at late times is $\propto t^{-1}$ \citep{LeitnerKravtsov_2011ApJ...734...48L}, resulting in $\epsilon_* \propto t^{-0.3}$.

Predictions for how the critical $\sigma_v$ for quenching should depend on redshift need to account for the fact that the zero-point for $t$ should coincide with the end of rapid star formation, which may not be the same for all galaxies in a given sample.  This paper will therefore not attempt a detailed analysis.  Instead, we present an illustrative example for a population of galaxies that formed most of their stars by $\sim 3$~Gyr after the big bang ($z \approx 2$).  One billion yr later ($z \approx 1.5$), most of the heat generated by the aging stellar population was from SN Ia.  The value of $\epsilon_*$ then declined by a factor of $\approx 2$ (because $\epsilon_* \propto t^{-0.3}$) during the ensuing 9~Gyr, implying that the $\sigma_v$ scale at which $\alpha_K \approx 2/3$ was a factor $\approx 1.4$ greater at $z \approx 1.5$ than at $z \approx 0$.

Optical observations suggest that the actual quenching scale has declined by a similar factor, but  quantitative comparisons await greater consistency among the definitions of quenching as a function of redshift.  For example, \citet{Franx_2008ApJ...688..770F} defined the stellar surface density threshold for quenching to be the level at which the specific star formation rate drops more than a factor of 3 below the prevailing rate at low stellar surface density.  They found that this surface density threshold was a factor $\sim 2.5$ smaller at $z \approx 0$ than at $z \approx 1.5$, corresponding to a factor of $\sim 1.6$ in $\sigma_v$.  A similar study by \citet{vanDokkum_2015ApJ...813...23V} focused more narrowly on the properties of massive compact galaxies at $1.5 < z < 2.25$, finding that their properties are consistent with a quenching probability that rises from zero at $\sigma_v = 220 \, {\rm km \, s^{-1}}$ to unity at $\sigma_v = 320 \, {\rm km \, s^{-1}}$.  When translated to low redshift through division by the predicted factor of 1.4, this range shifts to $160 \, {\rm km \, s^{-1}} \lesssim \sigma_v \lesssim 230 \, {\rm km \, s^{-1}}$, consistent with the low-redshift model outlined in this paper.  \citet{Whitaker_2017ApJ...838...19W} focused instead on the interval $0.5< z < 2.5$ and examined multiple definitions of quenching.  Their work aligns with \citet{vanDokkum_2015ApJ...813...23V} at $z = 2$ and is consistent with the central density threshold decline seen by \citet{Franx_2008ApJ...688..770F} down to $z = 0.5$.

\subsection{Quenching in Galaxy Cluster Cores}
\label{sec-ClusterCores}

We do not expect star-formation quenching in the central galaxies of galaxy clusters to depend as directly on $\sigma_v$, because their potential wells are much deeper than the stellar velocity dispersion of the central galaxy would indicate.  In the generic galaxy model of \S \ref{sec-GenericGalaxy}, we assumed that the maximum circular velocity of the halo around a massive galaxy was similar to that of the galaxy itself.  That assumption applies to galaxy groups with a velocity dispersion of $\sim 300 \, {\rm km \, s^{-1}}$ and an X-ray temperature of $\sim 1$~keV but not to galaxy groups and clusters with a velocity dispersion $\gtrsim 450 \, {\rm km \, s^{-1}}$ and an X-ray temperature $\gtrsim 2$~keV.  

Because of the deeper halo potential well, the gas pressure in the central galaxy of a galaxy cluster is often considerably greater than in the galaxies this paper has analyzed, and so radiative cooling can greatly exceed SN Ia heating.  Cool-core clusters, with a central entropy level $< 30 \, {\rm keV \, cm^2}$ and $t_{\rm cool} \lesssim 1$~Gyr at 10~kpc, have the largest central gas pressure.  Observations show that extended multiphase gas is nearly always present near the center of a cool-core cluster, independent of $\sigma_v$.  The AGN feedback limits the cooling flows in those central galaxies but allows multiphase gas to collect and star formation to proceed at 1\%--10\% of the uncompensated cooling-flow rate \citep{McDonald_2018ApJ...858...45M}.  In that regard, central galaxies in cool-core clusters are similar to the galaxies in Figure~\ref{schematic_K_plot_data_steady} with $200 \, {\rm km \, s^{-1}} < \sigma_v < 240 \, {\rm km \, s^{-1}}$.

Some of those central cluster galaxies have $\sigma_v > 240 \, {\rm km \, s^{-1}}$, meaning that there is a halo mass above which the mechanism outlined in \S \ref{sec-TheValve} does not work as described. In order for the mechanism to operate, the central AGN must be capable of lowering the CGM pressure by pushing much of the ambient halo gas to greater altitudes.  For that to happen, the total amount of kinetic feedback energy must be at least as great as the CGM binding energy, which is several times $10^{62} \, {\rm erg}$ in a $10^{14} \, M_\odot$ halo.  Significant lifting of that CGM requires $\sim 10^{45} \, {\rm erg \, s^{-1}}$ of energy input, sustained over a cosmological timescale.  The analogous requirement for a $10^{15} \, M_\odot$ halo is several times $10^{46} \, {\rm erg \, s^{-1}}$.  Observations of feedback in galaxy clusters show that the AGN's kinetic power is rarely greater than a few times $10^{45} \, {\rm erg \, s^{-1}}$ \citep[e.g.,][]{McNamaraNulsen2012NJPh...14e5023M}, accounting for why essentially all cool-core clusters with AGN feedback also have extended multiphase gas, regardless of $\sigma_v$ in the central galaxy.  However, why kinetic AGN power in the centers of galaxy clusters is not greater remains an open question.

\subsection{Quenching and the $M_{\rm BH}$--$\sigma_v$ Relation}
\label{sec-MBH_sigmav}

The black hole feedback valve mechanism for quenching implies that $M_{\rm BH}$ should depend on $\sigma_v$ because it requires the black hole's integrated kinetic energy output to be at least comparable to the binding energy of the CGM.  The resulting lower limit can be expressed as
\begin{equation}
  M_{\rm BH} \: \gtrsim \: \frac {f_{\rm b} M_{200} \sigma_v^2} {\epsilon_{\rm BH} c^2} 
             \:  \approx \: 10^8 \, M_\odot \left( \frac {\epsilon_{\rm BH}} {10^{-2}} \right)^{-1} 
             \sigma_{240}^5 
             \label{eq-MBH}
  \; \; ,
\end{equation}
where $\epsilon_{\rm BH}$ is the proportion of the black hole's rest-mass energy that becomes thermalized in the CGM.  A similar limit follows from setting $\epsilon_{\rm BH} M_{\rm BH} c^2 \gtrsim L_{\rm X} H_0^{-1}$ \citep{Voit_PrecipReg_2015ApJ...808L..30V}.  These limits are generic to any quenching model that relies primarily on AGN feedback to reduce the CGM pressure around a central galaxy \citep[e.g.,][]{Davies_2019MNRAS.485.3783D,Davies_2020,Oppenheimer_2020}, but the efficiency factor $\epsilon_{\rm BH}$ is difficult to infer from simple models.  Observations showing that $M_{\rm BH} \approx 7 \times 10^8 \, \sigma_{240}^{4.4} \, M_\odot$ \citep[e.g.,][]{KormendyHo2013ARAA..51..511K} suggest that the scaling of $M_{\rm BH}$ with $\sigma_v$ may result from the CGM lifting requirement of AGN feedback.  And the observations still imply $\epsilon_{\rm BH} \gtrsim 1.4 \times 10^{-3} \, \sigma_{240}^{-0.6}$ even if the connection is not directly causal.

In cosmological simulations, the conversion efficiency $\epsilon_{\rm kin}$ of accreted rest-mass energy to kinetic AGN feedback output is generally a free parameter. \citet{Oppenheimer_2020} recently analyzed an EAGLE simulation in which that conversion efficiency was set to $\epsilon_{\rm kin} = 1.67 \times 10^{-2}$.  They found that star-formation quenching was strongly correlated with both $M_{\rm BH}$ and lifting of the CGM.  They also found that CGM lifting is inefficient, because the integrated AGN feedback output required to accomplish CGM lifting was $\approx 10$ times the CGM binding energy.  Their simulations therefore effectively had $\epsilon_{\rm BH} \approx 1.7 \times 10^{-3}$, in good agreement with the efficiency inferred for CGM lifting from the $M_{\rm BH}$--$\sigma_v$ relation.  While that finding is encouraging, it cannot be considered clinching evidence in favor of a connection between the $M_{\rm BH}$--$\sigma_v$ relation and CGM lifting, because the effective $\epsilon_{\rm BH}$ still depends linearly on the arbitrarily tunable parameter $\epsilon_{\rm kin}$.  

Nevertheless, recent observations also indicate a close connection between quenching and integrated AGN energy output, as reflected by $M_{\rm BH}$.  \citet{Terrazas_2016ApJ...830L..12T,Terrazas_2017ApJ...844..170T} compared star formation rates among galaxies with directly measured central black hole masses and find a strong correlation between $M_{\rm BH}$ and quenching.  In quenched galaxies, the black hole masses are an order of magnitude greater than in star-forming galaxies of similar stellar mass.  The correlation between quenching and $\sigma_v$ in their sample is equivalently strong, with the transition to a quenched state occurring in the range $160 \, {\rm km \, s^{-1}} < \sigma_v < 250 \, {\rm km \, s^{-1}}$.  \citet{Terrazas_2016ApJ...830L..12T,Terrazas_2017ApJ...844..170T} found considerably weaker correlations between quenching and either $M_*$ or bulge mass.

According to the black hole feedback valve model, the galaxy property most fundamentally related to quenching is $\sigma_v$ because it determines (1) how effectively AGN feedback can suppress condensation of hot CGM gas and (2) the amount of AGN feedback energy necessary to lift the CGM out of the galactic potential well.  If that interpretation is correct, then the scatter observed in the  $M_{\rm BH}$--$\sigma_v$ relation primarily reflects the scatter in the efficiency parameter $\epsilon_{\rm BH}$ from galaxy to galaxy, at least among galaxies that are not at the centers of massive galaxy clusters.  However, the integrated AGN energy input needed to offset radiative cooling in the most massive halos is greater than one would infer from equation (\ref{eq-MBH}), causing the model's predictions to shift to greater $M_{\rm BH}$ at fixed $\sigma_v$ \citep[see][]{Voit_PrecipReg_2015ApJ...808L..30V}.

\subsection{Quenching and CGM Stripping}
\label{sec-Stripping}

Reducing the boundary pressure around a massive galaxy by stripping its CGM will have effects on quenching similar to those of AGN feedback.  Figure~\ref{circflow_K_plot_models} shows that SN Ia heating in galaxies with $\sigma_v \lesssim 200 \, {\rm km \, s^{-1}}$ is capable of driving a heated outflow that escapes the galaxy's potential well once CGM confinement has become negligible.  Those flows have $\min( t _{\rm cool} / t_{\rm ff} ) \gg 10$ and strong entropy gradients, implying that they remain homogeneous while escaping the galaxy. They also reach $t_{\rm cool} > 10 \, {\rm Gyr}$ inside of $r = 10 \, {\rm kpc}$.   Any cold gas remaining within the galaxy after the CGM is stripped may continue to form stars but will not be replenished by multiphase condensation of the CGM.  However, the outflow is probably too diffuse to prevent accretion of cosmological gas that may happen to be falling into the galaxy.

\subsection{Angular Momentum}
\label{sec-AngMom}

So far, our model completely ignores the undoubtedly important role of angular momentum.  In general, rotation is expected to promote development of a multiphase medium by suppressing the stabilizing effects of buoyancy on condensation \citep{Gaspari_2015A&A...579A..62G,SobacchiSormani_2019MNRAS.486..205S,SormaniSobacchi_2019MNRAS.486..215S}.  A rotating CGM is therefore more likely to condense at greater levels of $\min ( t_{\rm cool} / t_{\rm ff} )$, making quenching less likely at a given $\sigma_v$.  We plan to explore the effects of angular momentum in future work. This limitation currently precludes us from applying the model directly to late-type galaxies, in which angular momentum will be more important.

\section{Summary}
\label{sec-Summary}

We have presented a model for AGN feedback that closely links quenching of star formation with a galaxy's central stellar velocity dispersion.  That link emerges from an analysis of steady gaseous outflows driven by quiescent stellar populations.  We demonstrate that an outflow's profiles of pressure, density, and gas entropy depend directly on $\epsilon_*/v_c^2$, the ratio of the specific energy of ejected stellar gas to the square of the galaxy's circular velocity.  Galaxies with $\sigma_v \gtrsim 240 \, {\rm km \, s^{-1}}$ can remain in a steady state consisting of an inner cooling flow surrounded by a supernova-heated outflow, while galaxies with $\sigma_v \lesssim 240 \, {\rm km \, s^{-1}}$ are prone to multiphase circulation.  The AGN feedback in the subset with $\sigma_v \gtrsim 240 \, {\rm km \, s^{-1}}$ is therefore able to tune itself to lift the CGM, reduce the confining pressure it exerts, and quench star formation through a mechanism we have called the black hole feedback valve.  In galaxies with $\sigma_v \lesssim 240 \, {\rm km \, s^{-1}}$, this quenching mechanism becomes less reliable because feedback is more likely to produce multiphase precipitation away from the galaxy's center.  Figure~\ref{vc_alphaK_composite} graphically summarizes the model.

\begin{figure*}[t]
\begin{center}
\includegraphics[width=7.0in, trim = 0.0in 0.0in 0.0in 0.0in]{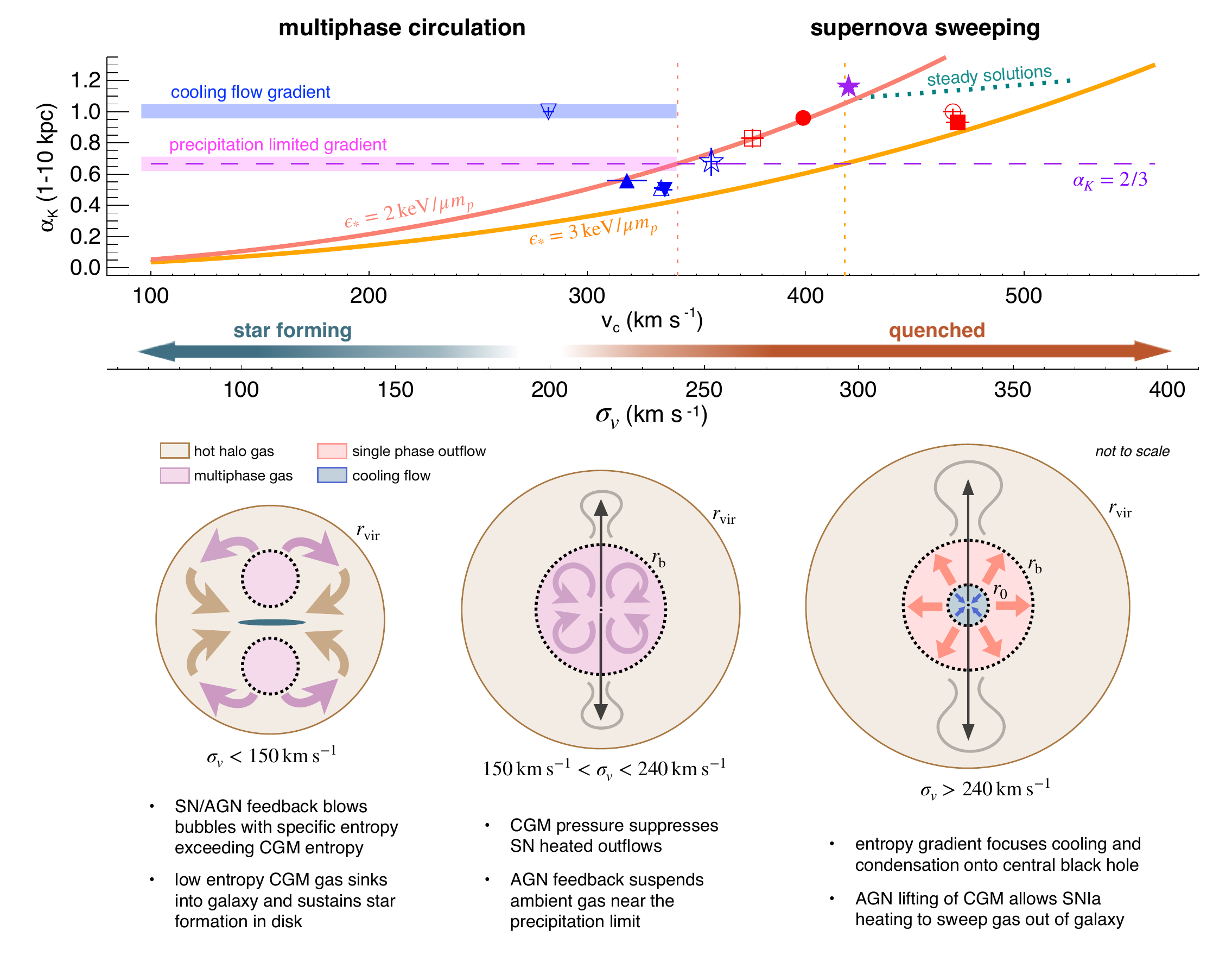} \\
\end{center}
\caption{ \footnotesize 
Graphical summary of the black hole feedback valve model.  At the top is a graph showing  relationships between the power-law entropy slope $\alpha_K$ of a galaxy's ambient gas and the circular velocity $v_c$ of its potential well.  The lower horizontal axis gives $\sigma_v = v_c / \sqrt{2}$, along with arrows indicating how galactic star formation correlates with $\sigma_v$ (\S \ref{sec-Dependence}).  A salmon colored line shows the predictions of equation (\ref{eq-alpha_K_vc}) for supernova-heated outflows driven by a specific stellar heat input, $\epsilon_* = 2 \, {\rm keV} / \mu m_p$.  An orange line shows predictions for $\epsilon_* = 3 \, {\rm keV} / \mu m_p$.  A dashed purple line shows the critical entropy slope $\alpha_K = 2/3$ (\S \ref{sec-CriticalSlope}).  Dotted salmon and orange lines indicate the values of $v_c$ that are critical for each value of $\sigma_v$.  A dotted teal line illustrates how the family of steady flow solutions shown in Figure~\ref{schematic_K_plot_models} departs from the predictions of equation (\ref{eq-alpha_K_vc}) above $\sigma_v \approx 300 \,{\rm km \, s^{-1}}$ and saturates near $\alpha_K \approx 1.2$.  Symbols correspond to the galaxies shown in Figure~\ref{Werner_equality_red_blue} and represent the best-fitting values of $\alpha_K$ from 1 to 10 kpc.  Within galaxies having $200 \, {\rm km \, s^{-1}} < \sigma_v < 240 \, {\rm km \, s^{-1}}$, radiative cooling exceeds supernova heating in this radial interval (\S \ref{sec-HC_Equality}), and so $\alpha_K$ is not expected to follow the predictions of equation (\ref{eq-alpha_K_vc}).  Most have entropy slopes similar to the precipitation limit (magenta line), but one has a slope closer to that of a pure cooling flow (blue line).  Below the graph are three schematic illustrations of the model's qualitative predictions for how the flow pattern in a galaxy's ambient medium should depend on $\sigma_v$.   On the left is a lower-mass galaxy around which feedback blows bubbles with greater specific entropy than the CGM.   Those buoyant bubbles drive multiphase circulation and fail to prevent cold, star-forming gas from collecting in the galaxy's disk (\S \ref{sec-Dependence}).  In the middle is a more massive galaxy in which CGM entropy exceeds what supernova heating can produce.  The CGM pressure therefore confines ejected stellar gas and causes some of it to accrete onto the central black hole.  Then, AGN feedback suspends the ambient medium in a marginally precipitating state, driving multiphase circulation but preventing significant star formation.  In the high-mass galaxy on the right, supernova heating beyond the stagnation radius $r_0$ can drive an outflow with $\alpha_K > 2/3$ because $\sigma_v > 240 \, {\rm km \, s^{-1}}$.  Cooling and condensation of ambient gas is therefore focused on the central black hole, which responds by producing strong jets that heat the CGM.  In this configuration, CGM pressure at the boundary radius $r_{\rm b}$ determines the strength of the cooling flow inside of $r_0$ and therefore acts like the knob on a valve that governs AGN feedback power.  The valve adjusts itself so that time-integrated AGN feedback power suffices to lift the much of the CGM out of the galaxy's potential well, leading to complete quenching of star formation.
\vspace*{1em}
\label{vc_alphaK_composite}}
\end{figure*}

The most notable features of the model are as follows.

\begin{enumerate}

\item It predicts that a radial supernova-heated outflow through a massive galaxy should have a power-law entropy profile slope ($\alpha_K$) that depends primarily on $\epsilon_*/v_c^2$ (\S \ref{sec-Sweeping}).

\item The entropy slope $\alpha_K \approx 2/3$ is special because it corresponds to an electron density profile $n_e \propto r^{-1}$ along which the ratio of SN Ia heating to radiative cooling remains approximately constant with radius.  It is also special because the $t_{\rm cool} / t_{\rm ff}$ ratio remains approximately constant with radius.  Galactic outflows that have $\alpha_K > 2/3$ therefore tend to promote cooling and condensation of ambient gas near the origin, in the vicinity of the central black hole, and become less prone to multiphase condensation as $r$ increases.  Conversely, outflows with $\alpha_K < 2/3$ are more prone to condensation at large radii and therefore promote multiphase circulation.  For a stellar population age of $\approx 10$~Gyr, the critical slope $\alpha_K \approx 2/3$ corresponds to $\sigma_v \approx 240 \, {\rm km \, s^{-1}}$ (\S \ref{sec-CriticalSlope}).

\item The X-ray observations of massive elliptical galaxies agree with the heated outflow analysis, in that galaxies with $\sigma_v > 240 \, {\rm km \, s^{-1}}$ tend to have SN Ia heating rates that exceed radiative cooling at $r \approx 1$--10~kpc, while radiative cooling tends to exceed SN Ia heating over the same radial range among galaxies with $\sigma_v < 240 \, {\rm km \, s^{-1}}$ (\S \ref{sec-HC_Equality}).

\item The normalizations of the density, pressure, and entropy profiles of a subsonic heated outflow are determined by an outer pressure boundary condition set by the CGM.  If the pressure is too large, radiative cooling will exceed SN Ia heating, causing a cooling flow that triggers AGN feedback.  Coupling between AGN feedback and CGM pressure should therefore keep the ambient gas near the AGN close to the point of heating/cooling balance (\S \ref{sec-BoundaryPressure}).

\item In galaxies with $\sigma_v \gtrsim 240 \, {\rm km \, s^{-1}}$, this coupling between AGN feedback and the CGM should form a self-regulating valve that links AGN fueling to CGM pressure.  Feedback in those galaxies adds heat to the CGM, causing it to expand until the reduction in its pressure brings time-averaged heating into balance with the energy needed to lift the CGM (\S \ref{sec-TheValve}).  This black hole feedback valve mechanism inevitably quenches star formation because the ambient value of $t_{\rm cool} / t_{\rm ff}$ rises beyond $\approx 20$ at $r \gtrsim 1$~kpc, making the ambient medium stable to multiphase condensation (\S \ref{sec-ThermalStability}).

\item In galaxies with $\sigma_v \lesssim 240 \, {\rm km \, s^{-1}}$, outflows heated by SN Ia and confined by a significant CGM pressure cannot remain heating-dominated as they propagate to large radii.  They are destined to become cooling-dominated, producing entropy inversions that are unstable to multiphase condensation.  Those galaxies are consequently prone to precipitation and evolve toward a precipitation-limited state with $t_{\rm cool} / t_{\rm ff} \approx 10$--20 over a broad range of radii (\S \ref{sec-ThermalStability}).

\item Simulations designed to test the black hole feedback valve mechanism require high spatial resolution ($< 100$~pc), because the mechanism calls for high-powered jets ($\gtrsim 10^{44} \, {\rm erg \, s^{-1}}$) to drill through the galaxy's ambient medium at 1--10~kpc without significantly disrupting it before thermalizing their energy in the CGM at 10--100~kpc (\S \ref{sec-JetPropagation}).

\item Numerical steady flow solutions corroborate the analytical estimates on which the black hole feedback valve model is based.  In particular, steady flows in galaxies with $\sigma_v \gtrsim 240 \, {\rm km \, s^{-1}}$ are cooling flows at small radii and SN Ia-heated outflows at larger radii, with $\alpha_K$ in alignment with the predictions of equation (\ref{eq-alpha_K_vc}).\footnote{The alignment is poorer above $\sigma_v \approx 300 \, {\rm km \, s^{-1}}$, because the mean entropy slope of the steady flow solutions at 1--10 kpc starts to saturate at $\alpha_K \approx 1.2$, as shown by the dotted teal line in Figure~\ref{vc_alphaK_composite}.}  Steady flows in galaxies with $\sigma_v \lesssim 240 \, {\rm km \, s^{-1}}$, on the other hand, are cooling dominated at large radii and are prone to developing entropy inversions (\S \ref{sec-SteadyFlow}).

\end{enumerate}

Comparisons with high-quality {\em Chandra} X-ray observations of 10 massive elliptical galaxies support the model. 

\begin{enumerate}

\item  In the galaxies with $\sigma_v \gtrsim 240 \, {\rm km \, s^{-1}}$, the entropy slope $\alpha_K$ from $\sim 1$ to 10~kpc generally agrees with the predictions of  equation (\ref{eq-alpha_K_vc}), and those galaxies are single phase in that radial interval, with one exception (NGC~6868, possibly a borderline case with $\sigma_v = 252 \, {\rm km \, s^{-1}}$).  The galaxies with $\sigma_v \lesssim 240 \, {\rm km \, s^{-1}}$, in contrast, are multiphase in that radial interval and track the precipitation limit at $t_{\rm cool} / t_{\rm ff} \approx 10$ (\S \ref{sec-ObsComparison}).   Outside of 10~kpc, ambient gas around the multiphase galaxies tends to have lower entropy, greater pressure, and greater density, resulting in greater X-ray luminosities relative to the single-phase galaxies.  Larger samples of elliptical galaxies should therefore be checked to see if there is an inflection of the $L_{\rm X}$-$\sigma_v$ relation above $240 \, {\rm km \, s^{-1}}$ once the central galaxies of groups and clusters with $kT \gtrsim 2$~keV have been excluded.

\item Inside of $\sim 0.5$~kpc, the entropy profiles of most of the ellipticals analyzed depart from the power laws observed at larger radii and flatten near $K_0 \approx 2 \, {\rm keV \, cm^2}$.  That entropy level is consistent with intermittent shock heating of a precipitation-limited atmosphere by $\sim 10^{42} \, {\rm erg \, s^{-1}}$ of kinetic feedback.  Bondi accretion of ambient gas at that entropy level onto a central black hole with $M_{\rm BH} \sim 10^9 \, M_\odot$ is capable of supplying the currently observed feedback power, but it is insufficient to lift the CGM.  However, the one galaxy with an unbroken power-law entropy distribution inside of 0.5~kpc (NGC~4261) has a kinetic power output 2 orders of magnitude greater.  It is also the only galaxy in the sample with $t_{\rm cool} / t_{\rm ff} \lesssim 10$ at $< 200$~pc, implying that chaotic cold accretion onto the central black hole is temporarily supercharging AGN feedback and enabling it to lift the CGM (\S \ref{sec-SinglePhase}, see also V15).

\end{enumerate}

The model was inspired by X-ray observations but has broad implications for optical/IR studies of galaxy evolution (\S \ref{sec-Implications}):

\begin{enumerate}

\item  It links the ability of AGN feedback to quench star formation directly to $\sigma_v$, which optical/IR observations have shown to be the galaxy attribute most closely correlated with quenching of central galaxies \citep{Wake_2012ApJ...751L..44W,Teimoorinia_2016MNRAS.457.2086T,Bluck_2016MNRAS.462.2559B,Bluck_2020MNRAS.492...96B}.  At $\sigma_v \gtrsim 240 \, {\rm km \, s^{-1}}$, observations show that $\gtrsim 90$\% of both central and satellite galaxies are quenched, with no apparent dependence on $\sigma_v$.  This finding aligns well with the model's prediction that supernova heating outflows should be homogeneous in galaxies with $\sigma_v \gtrsim 240 \, {\rm km \, s^{-1}}$.  In contrast, the fraction of galaxies with $\sigma_v \lesssim 240 \, {\rm km \, s^{-1}}$ that are quenched appears to be a continuous function of $\sigma_v$ that drops below 25\% for central galaxies with $\sigma_v \lesssim 100 \, {\rm km \, s^{-1}}$.  According to the model, the quenched fraction should be smaller at smaller $\sigma_v$ because energetic central feedback more easily produces entropy inversions that result in multiphase condensation (\S \ref{sec-Dependence}).  However, this qualitative conclusion needs to be investigated more quantitatively with numerical simulations.

\item Earlier in time, the predicted critical value of $\sigma_v$ for quenching is greater, because $\epsilon_*$ is greater in younger stellar populations.  While star formation is still active, the critical value for effective quenching by AGN feedback is $\sigma_v \approx 300 \, \eta^{-1} \, {\rm km \, s^{-1}}$, where $\eta$ is the ratio of gas outflow rate to star formation rate.  It implies that galaxies cannot greatly exceed $\sigma_v \approx 300 \, {\rm km \, s^{-1}}$ without supercharging AGN feedback. The properties of ``red nugget" galaxies at $z \sim 2$ support this implication because they are compact, quenched galaxies with $250 \, {\rm km \, s^{-1}} \lesssim \sigma_v \lesssim 400 \, {\rm km \, s^{-1}}$ (\S \ref{sec-UpperLimit}).

\item As a galaxy's stellar population ages, the critical value of $\sigma_v$ declines because the specific SN Ia rate drops more rapidly than the specific stellar mass-loss rate.  This decline plausibly accounts for observations showing that the central stellar surface density associated with quenching declines with time (\S \ref{sec-Evolution}).

\item In the central galaxies of massive galaxy clusters, $\alpha_K$ is not expected to correlate as closely with $\sigma_v$, because the greater central CGM pressure in cool-core clusters causes radiative cooling to greatly exceed SN Ia heating.  Those systems are instead observed to have roughly constant $t_{\rm cool} / t_{\rm ff}$ profiles that track the precipitation limit, with $\alpha_K \approx 2/3$ at 5--20~kpc (\S \ref{sec-ClusterCores}).

\item Any model for quenching that requires AGN feedback to alleviate CGM pressure confinement, including the one presented in this paper, predicts a lower limit on $M_{\rm BH}$ that depends on $\sigma_v$.  The resulting scaling ($M_{\rm BH} \propto \sigma_v^5$) is interestingly close to the observed scaling.  If central black hole mass is indeed linked to the binding energy of the CGM, then observations imply that AGN feedback thermalizes a fraction $\gtrsim 10^{-3}$ of the black hole's rest-mass energy in the surrounding CGM (\S \ref{sec-MBH_sigmav}).

\item Galaxies in which rotation helps to support the ambient medium might not conform as precisely to the model's predictions, because of how angular momentum alters the buoyancy effects that limit condensation.  Werner et al.~(2014) have suggested that NGC 6868 may be one such example \citep[see also][]{Juranova_2019MNRAS.484.2886J}.  Future modeling will therefore need to account for how rotation affects the critical value of $\sigma_v$ (\S \ref{sec-AngMom}).

\end{enumerate}

\vspace*{2em} 

G.M.V. acknowledges helpful conversations with Iu. Babyk, A. Bluck, L. Ciotti, A. Evrard, D. Maoz, B. McNamara, B. Oppenheimer, and B. Terrazas and support from Chandra Science Center grant TM8-19006X.  G.B. acknowledges support from NSF (grant AST-1615955, OAC-1835509), and NASA (grant NNX15AB20G), and computational support from NSF XSEDE.  B.W.O. acknowledges support from the NSF (AST-1517908), NASA ATP (15AP39G), and 80NSSC18K1105


\bibliographystyle{apj}

\end{document}